\documentclass[aps,prb,twocolumn,superscriptaddress, amsmath]{revtex4-2}

\usepackage[]{graphicx}

\newcommand{\upmu}{\relax{\mbox{\usefont{U}{psy}{m}{n}{m}}}}

\newcommand{\tg}{$t_{2g}$}

\newcommand{\dxy}{$d_{xy}$}
\newcommand{\dxz}{$d_{xz}$}

\newcommand{\mub}{$\mu_{\rm B}$}

\usepackage[normalem]{ulem}
\usepackage{color}
\begin{document}


\title{Angular dependence of Hall effect and magnetoresistance in SrRuO$_3$-SrIrO$_3$ heterostructures}

\author{Sven Esser}
\affiliation{Experimentalphysik VI, Center for Electronic Correlations and Magnetism, Augsburg University, D-86159 Augsburg, Germany}

\author{Jiongyao Wu}
\affiliation{Department of Physics and Center for Nanointegration Duisburg-Essen (CENIDE), University of Duisburg-Essen, Lotharstr. 1, 47057 Duisburg, Germany}

\author{Sebastian Esser}
\author{Robert Gruhl}
\author{Anton Jesche}

\affiliation{Experimentalphysik VI, Center for Electronic Correlations and Magnetism, Augsburg University, D-86159 Augsburg, Germany}

\author{Vladimir Roddatis}
\affiliation{Institut f\"{u}r Materialphysik, Georg-August-Universit\"{a}t G\"{o}ttingen, D-37077 G\"{o}ttingen, Germany}
\affiliation{GFZ German Research Centre for Geosciences, Telegrafenberg 14473, Potsdam, Germany}

\author{Vasily Moshnyaga}
\affiliation{I. Physikalisches Institut, Georg-August-Universit\"{a}t G\"{o}ttingen, D-37077 G\"{o}ttingen, Germany}

\author{Rossitza Pentcheva}
\affiliation{Department of Physics and Center for Nanointegration Duisburg-Essen (CENIDE), University of Duisburg-Essen, Lotharstr. 1, 47057 Duisburg, Germany}

\author{Philipp Gegenwart}
\affiliation{Experimentalphysik VI, Center for Electronic Correlations and Magnetism, Augsburg University, D-86159 Augsburg, Germany}

\date{\today}

\begin{abstract}
	Perovskite SrRuO$_3$ is a prototypical itinerant ferromagnet which allows interface engineering of its electronic and magnetic properties. We report synthesis and investigation of atomically flat artificial multilayers of SrRuO$_3$ with the spin-orbit semimetal SrIrO$_3$ in combination with band-structure calculations with a Hubbard $U$ term and topological analysis. They reveal an electronic reconstruction and emergence of flat Ru-4d$_{xz}$ bands near the interface, ferromagnetic interlayer coupling and negative Berry-curvature contribution to the anomalous Hall effect. We analyze the Hall effect and magnetoresistance measurements as a function of the field angle from out of plane towards in-plane orientation (either parallel or perpendicular to the current direction) by a two-channel model. The magnetic easy direction is tilted by about $20^\circ$ from the sample normal for low magnetic fields, rotating towards the out-of-plane direction by increasing fields. Fully strained epitaxial growth enables a strong anisotropy of magnetoresistance. An additional Hall effect contribution, not accounted for by the two-channel model is compatible with stable 
	skyrmions only up to a critical angle of roughly $45^\circ$ from the sample normal. Within about $20^\circ$ from the thin film plane an additional peak-like contribution to the Hall effect suggests the formation of a non-trivial spin structure. 
\end{abstract}


\maketitle

\section{Introduction}
The demand of high density and energy efficient electronic devices generated enormous interest on magnetic skyrmions, i.e., topologically protected spin textures~\cite{Skyrme1962}, during the last decade~\cite{Fert2017, Finocchio2016}. For instance, controlled modification of skyrmion information may be utilized in energy-efficient racetrack memory devices~\cite{Wang2019}. Magnetic skyrmions can be generated by different mechanisms. For thin films a promising strategy is the combination of a ferromagnetic (FM) material with a strong spin-orbit coupled (SOC) heavy metal, introducing a Dzyaloshinskii-Moriya interaction (DMI) at the interface due to the broken inversion symmetry~\cite{Nagaosa2013}.

Sharp inversion symmetry breaking interfaces can be grown in heterostructures of different perovskite oxides. Recently, Li {\it et al.}~\cite{Li2019} showed in bilayers of FM La$_{0.7}$Sr$_{0.3}$MnO$_3$ (LSMO) with the SOC semimetal SrIrO$_3$ (SIO) \cite{Nan2019, Fruchter2016} the appearance of a giant topological Hall effect (THE), related to the LSMO/SIO interface. In thin film heterostructures composed of the FM metal SrRuO$_3$ (SRO) \cite{Agrestini2015, Ishigami2015} and the ferroelectric (FE) BaTiO$_3$ (BTO), Wang {\it et al.}~\cite{Wang2018} found a broken inversion symmetry at the interface due to the FE proximity effect, enabling electric control of the skyrmion properties.

The combination of SRO and SIO grown on SrTiO$_3$ (STO) substrate offers an excellent possibility for growing epitaxial thin heterostructures with multiple interfaces. Both components crystallize in the same $ABO_3$ related perovskite structure as substrate STO with similar valence on both the $A$- and the $B$-sites, avoiding charge accumulation and polar catastrophe at the interface~\cite{Nakagawa2006}. Furthermore, their (pseudo)cubic lattice constants are rather similar, i.e., the lattice mismatch is smaller than $0.97\%$. 

Matsuno et al. \cite{Matsuno2016} reported first hints at the presence of skyrmions in bilayers of 2 unit-cell (u.c.) SIO on 4 to 7 u.c. SRO (on STO substrate) by analyzing an unconventional contribution to the Hall effect at temperatures below 100 K, which they interpreted as THE. Resulting from a Berry phase in real-space \cite{Everschor-Sitte2014}, THE is a promising indicator of skyrmions, first demonstrated in MnSi~\cite{Neubauer2009}.

For the inverted heterostructure, i.e., SRO (3 to 5 u.c.)-SIO (2 u.c.) on STO the AHE and THE contributions were successfully varied by electric field tuning and the change of the THE was attributed to a change in size of the topological spin textures by modification of the DMI~\cite{Ohuchi2018}. Similar THE was also observed in other studies on SRO-SIO superlattices (SLs) as well as in SRO films~\cite{Sohn2018, Wang2020a} and linked with the appearance of skyrmions~\cite{Pang2017, Meng2019}. By contrast, other groups related the anomalies in the Hall signal not to a topological contribution. Instead, they considered anomalous Hall resistance due to different magnetic domains in SRO \cite{Kan2018, Miao2020, Wang2020b}, the variation of the magnetic and electronic properties between individual SRO layers \cite{Yang2021}, thickness inhomogeneities in the SRO layers \cite{Wang2020} or defect-induced anomalies \cite{Kan2018a}. From tight-binding calculations Groenendijk et al. \cite{Groenendijk2018}  arranged the electronic structure of SRO in groups of 3 bands, with 2 low-lying non-trivial bands with Chern number $C=\pm2$. They describe the AHE effect phenomenologically by superposition of two channels arising from nontrivial Berry curvature in reciprocal space, although real-space Berry
curvature effects due to noncollinear spin textures could not be excluded. Thus, further experimental characterization of the Hall effect in SRO-SIO systems is strongly needed to clarify the situation.

Below, we report a detailed combined study of DFT+U+SO band-structure calculations and angular dependent magnetotransport and Hall effect on [(SRO)$_5$/(SIO)$_2$)]$_k$ (SROSIO)$_k$ ($k=1-10$) multilayer thin films. The latter were prepared by metalorganic aerosol deposition (MAD), which realizes an oxygen-rich growth atmosphere, ideal for high-quality perovskite oxide thin films \cite{Moshnyaga1999, Schneider2010, Schneider2014}. The band structure of the superlattice displays dispersive and emergent flat bands near the interface. Analysis of the experimental data with a phenomenological two-band model successfully explains the overall transport properties. An additional Hall contribution can be associated to skyrmions only when the field is applied at an angle between 0 and 45$^\circ$ from the sample normal.

\section{Methods}

Thin films of (SROSIO)$_k$ heterostructures were grown on (001) oriented STO substrates by MAD. The growth of each layer was monitored in-situ by optical ellipsometry which allows to resolve variations on a monolayer length scale~\cite{Jungbauer2014}. 
Film thickness was determined by X-ray reflectometry (XRR) utilizing a \texttt{Malvern Panalytical Empyrean} diffractometer and subsequently simulated with \texttt{ReMagX} \cite{Macke2014} software. Thereby the thicknesses of each layer and the interface roughness could be extracted. Crystal structure, phase purity and strain state were determined by X-ray diffraction (XRD) and reciprocal space mapping (RSM) operating with Cu-K$_{\alpha1}$ radiation by means of a hybrid monochromator.

The temperature- and magnetic field-dependent magnetization was measured in a \texttt{Magnetic Property Measurement System} (MPMS3) equipped with a 7\,T magnet in a stabilized dc mode. To obtain the proper magnetization of the thin film the raw background signal of the substrate material was carefully subtracted in each measurement point from the raw measurement signal of the sample and analyzed afterwards by the standard procedure of the MPMS. The temperature- and magnetic field-dependence of the electrical resistivity was determined within a \texttt{Physical} \texttt{Property} \texttt{Measurement} \texttt{System} (PPMS) equipped with a 14\,T magnet and an electrical transport option. Each thin film was micro-structured with argon-ion etching in a standard Hall bar geometry. The angle-dependent magneto-resistance and Hall-effect studies were performed with a PPMS horizontal rotator.

Density functional theory (DFT)~\cite{Sham} calculations were performed with the VASP~\cite{Frthmuller} code with the PBEsol~\cite{Burke, Perdew, PerdewJ} exchange-correlation functional, which is known to improve the structural description of solids. Static local correlation effects were considered within the DFT+$U$ approach in the Dudarev implementation~\cite{Dudarev} , using $U$ = 1.0 eV for Ru \cite{Spaldin}. For Ir we used  $U$ = 1.37 eV, $J$ = 0.22 eV, obtained from linear response theory for the bulk compound~\cite{Franchini}. We model the (SrRuO$_3$)$_n$/(SrIrO$_3$)$_2$(001) SL by using laterally a $\sqrt{2}a \times \sqrt{2}a$ unit cell, with $a$ set to the lattice constant of STO and two transition metal sites per layer to account for the octahedral rotations and tilts. Since an odd number of layers constrains the tilt pattern and the numerical demand for doubling the unit cell in $z$ direction for $n=5$ is too high, we have performed additional calculations with $n=4$ and for a [(SrRuO$_3$)$_3$/(SrIrO$_3$)$_2$]$_2$(001) SL. Since the results for $n=4$ and 5 were qualitatively very similar, we proceed here with the results for $n=5$ that corresponds to the experimental setup.  A plane-wave cutoff energy of 500 eV was used. The calculations were performed with a $\Gamma$ centered $k$-point grid of $6\times 6\times\ 2$. The convergence criterion of the self-consistent calculation is 10$^{-5}$ eV for total-energy. A Fermi level smearing of 0.05 eV is used for the geometry optimization and further analysis. The lattice parameter $c$ and the internal coordinates of the structure were optimized until the atomic forces were less than 0.01 eV/\AA. Spin-orbit coupling (SOC) was considered with magnetization direction along the [001] and [100] quantization axes, below we report the results for the [001] orientation. The  DFT+$U$+SOC band structure was fitted to maximally localized Wannier functions using the wannier90 code~\cite{Marzari} and the Berry curvature and anomalous Hall conductivity (AHC) were calculated  on a dense  $k$-point mesh of $100\times 100\times\ 10$.

\section{Results and Discussion}
\subsection{Structural investigations}
\begin{figure}[b!]
	\includegraphics[width=0.48\textwidth]{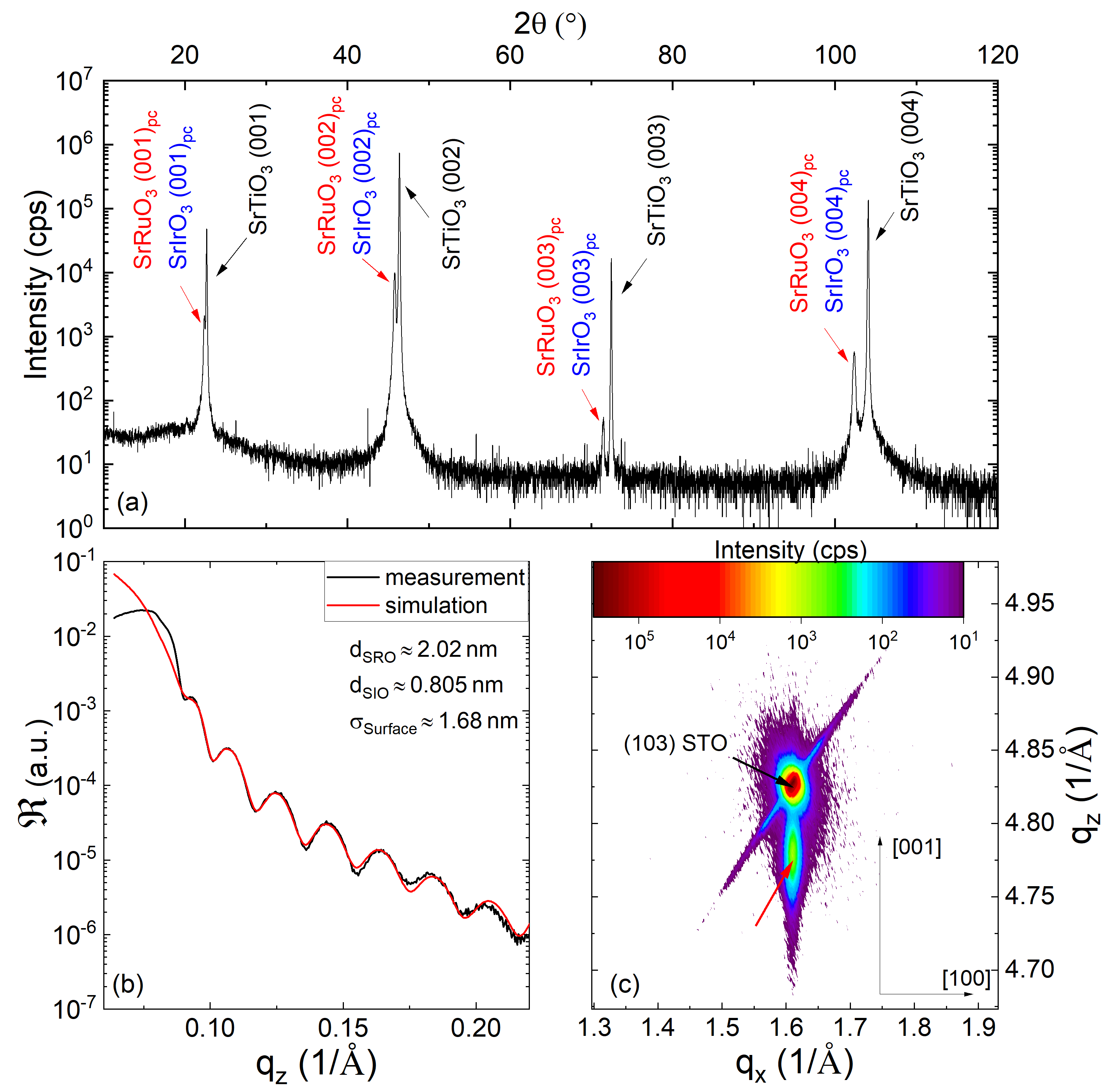}%
	\caption{\label{fig-XRD-XRR-RSM}(a) $\theta$--$2\theta$ XRD scan: Black arrows indicating peaks of the (001)-STO substrate, whereas red arrows mark peaks of (001)$_{\text{pc}}$-SRO and (001)$_{\text{pc}}$-SIO. (b) XRR measurement of a [(SRO)$_5$/(SIO)$_2$)]$_{10}$ heterostructure including the simulation with \texttt{ReMagX} \cite{Macke2014}. (c) Reciprocal space map around the (103) peak of STO (black arrow) indicating the fully strained state of the heterostructure marked by a red arrow.}
\end{figure}

\begin{figure}[b!]
	\includegraphics[width=0.48\textwidth]{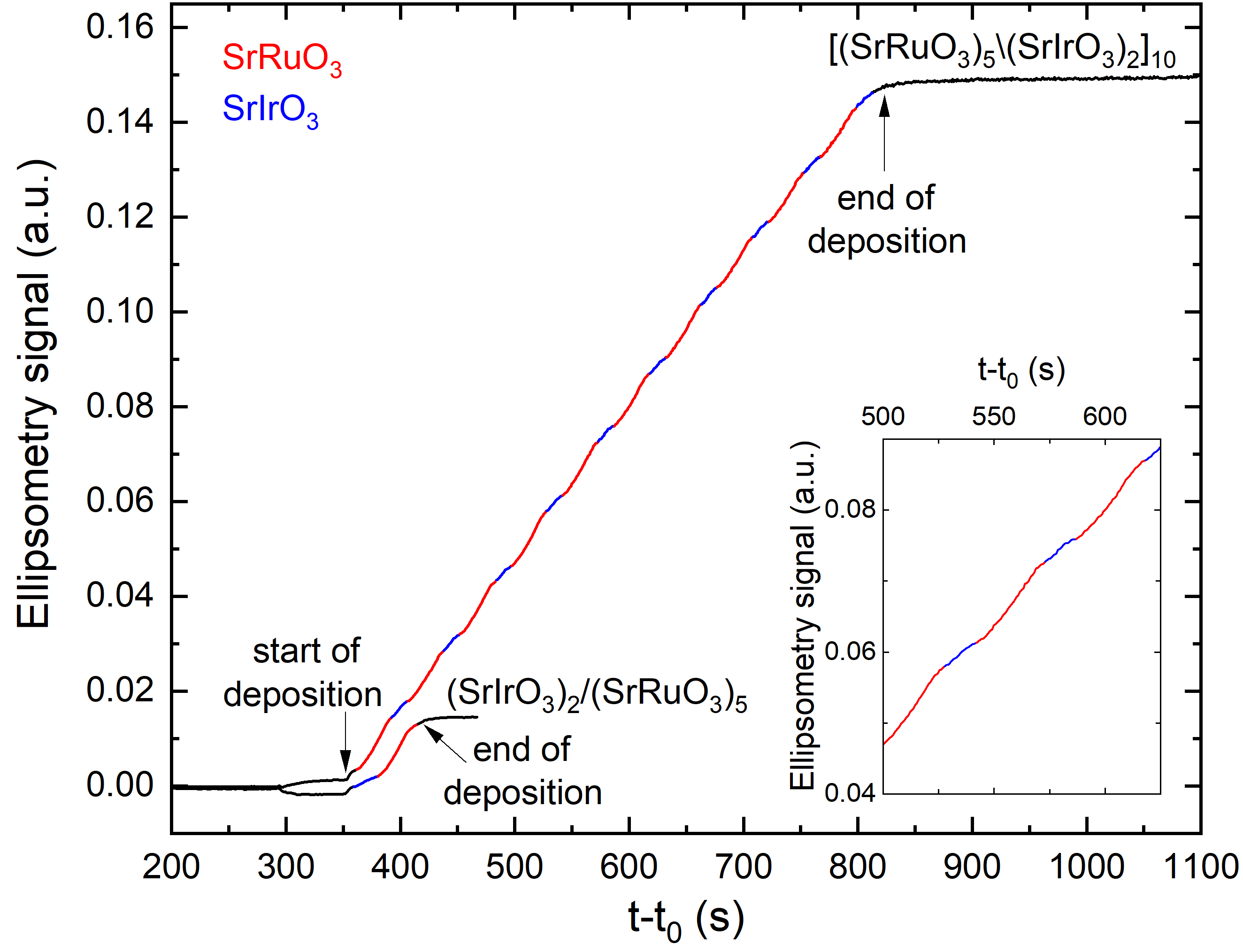}%
	\caption{\label{fig-Ellipsometry}Ellipsometry signal as function of time during the growth of a (SrIrO$_3$)$_2$/(SrRuO$_3$)$_5$ bilayer structure and a 10 times repeated bilayer structure ($[$(SrRuO$_3$)$_5$/(SrIrO$_3$)$_2]_{10}$). Black arrows indicate the start and the end of the deposition. The measured signals during the growth of SRO and SIO are colored in red and blue, respectively.}
\end{figure}

XRD measurements (Fig. \ref{fig-XRD-XRR-RSM}(a)) on (001)$_{\text{pc}}$ oriented SROSIO thin films on (001) oriented STO substrates indicate an out-of-plane epitaxial growth. The calculated pseudo-cubic out-of-plane lattice constant $d_{\text{pc}}=3.950(4)\,\mathring{\text{A}}$ is slightly larger than that of bulk SRO ($d_{\text{SRO}}=3.923\,\mathring{\text{A}}$) \cite{Jones1989} and SIO ($d_{\text{SIO}}=3.942\,\mathring{\text{A}}$) \cite{Zhao2008}. An explanation for this can be found in the compressive in-plane strain of $\sim0.44\%$ and $\sim0.97\%$ due to the lattice mismatch for SRO/STO and SIO/STO, respectively. Reciprocal space mapping around the (103)-STO peak (Fig. \ref{fig-XRD-XRR-RSM}(c)) confirms the fully strained state of the film. The occurrence of Laue fringes in the XRD pattern (see Supplemental Material \cite{Supplemental}) indicates the high quality of our films. Small-angle X-ray reflectometry measurements indicate a large-scale homogeneity for the growth of each SRO and SIO layer independently. This homogeneous growth is in good agreement with the measured time dependence of the imaginary part of the reflection coefficients during the synthesis, plotted in Fig. \ref{fig-Ellipsometry}. As shown previously,~\cite{Jungbauer2014,Azzam1975}, the phase shift detected by ellipsometry grows for ultrathin films of thickness smaller than the wavelength linearly with thickness, with a slope that depends on the materials optical constants and thus changes between the different layers. Thus, the heterostructure growth is clearly visible the time dependent signal.

\subsection{Band-structure calculations}

\begin{figure}[!htp]
	\includegraphics[width=0.5\textwidth]{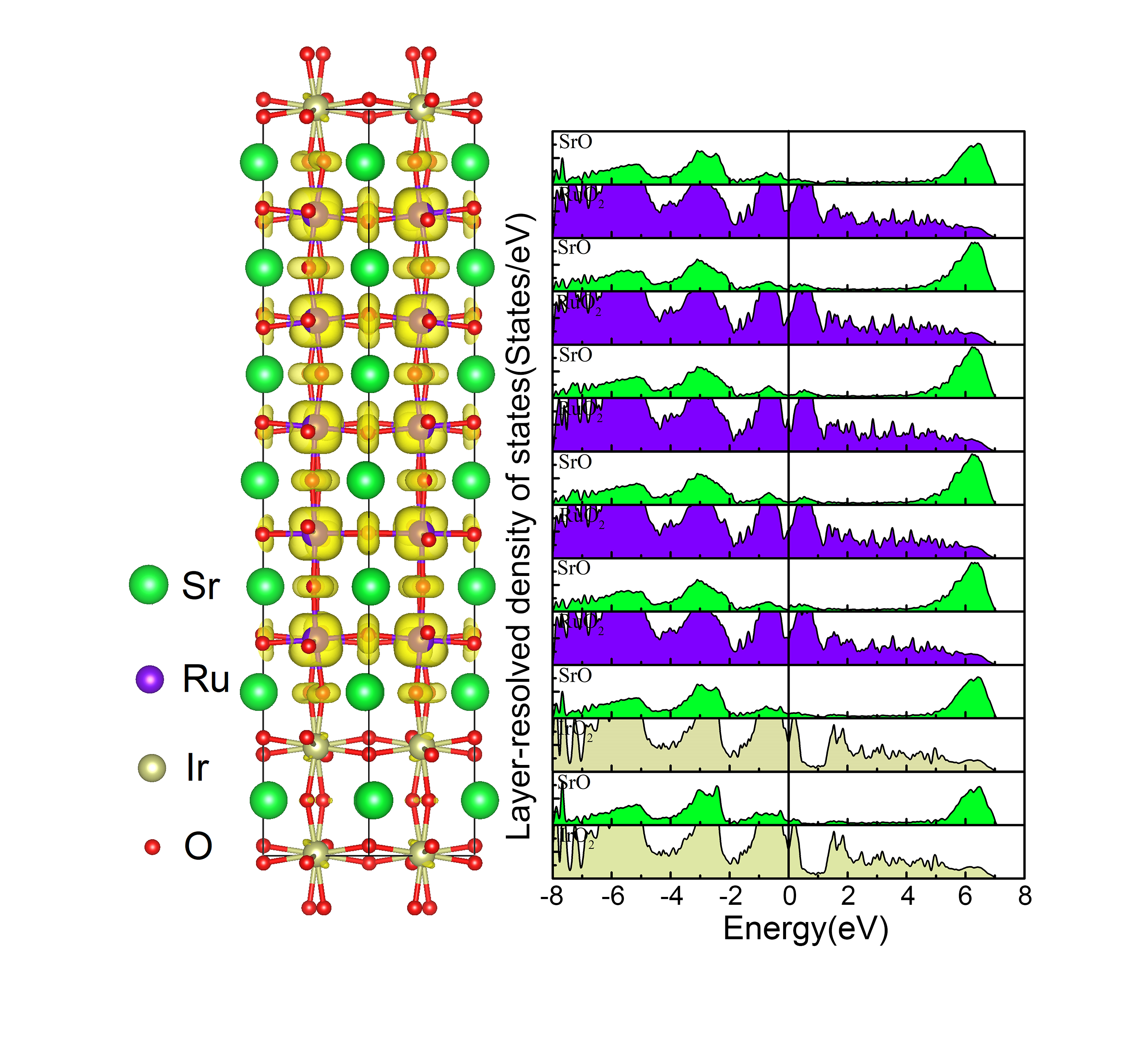}%
	\caption{\label{DFT_structure}Side view of the spin density (left) and layer-resolved density of states (LDOS, right) of the (SrRuO$_3$)$_5$/(SrIrO$_3$)$_2$(001) SL with SOC along [001]. Zero energy in the LDOS denotes the Fermi level. Note, that the magnetic coupling between the SRO blocks along the z-direction is ferromagnetic, see text.}
\end{figure}


\begin{figure*}[!htp]
	\includegraphics[width=1\textwidth]{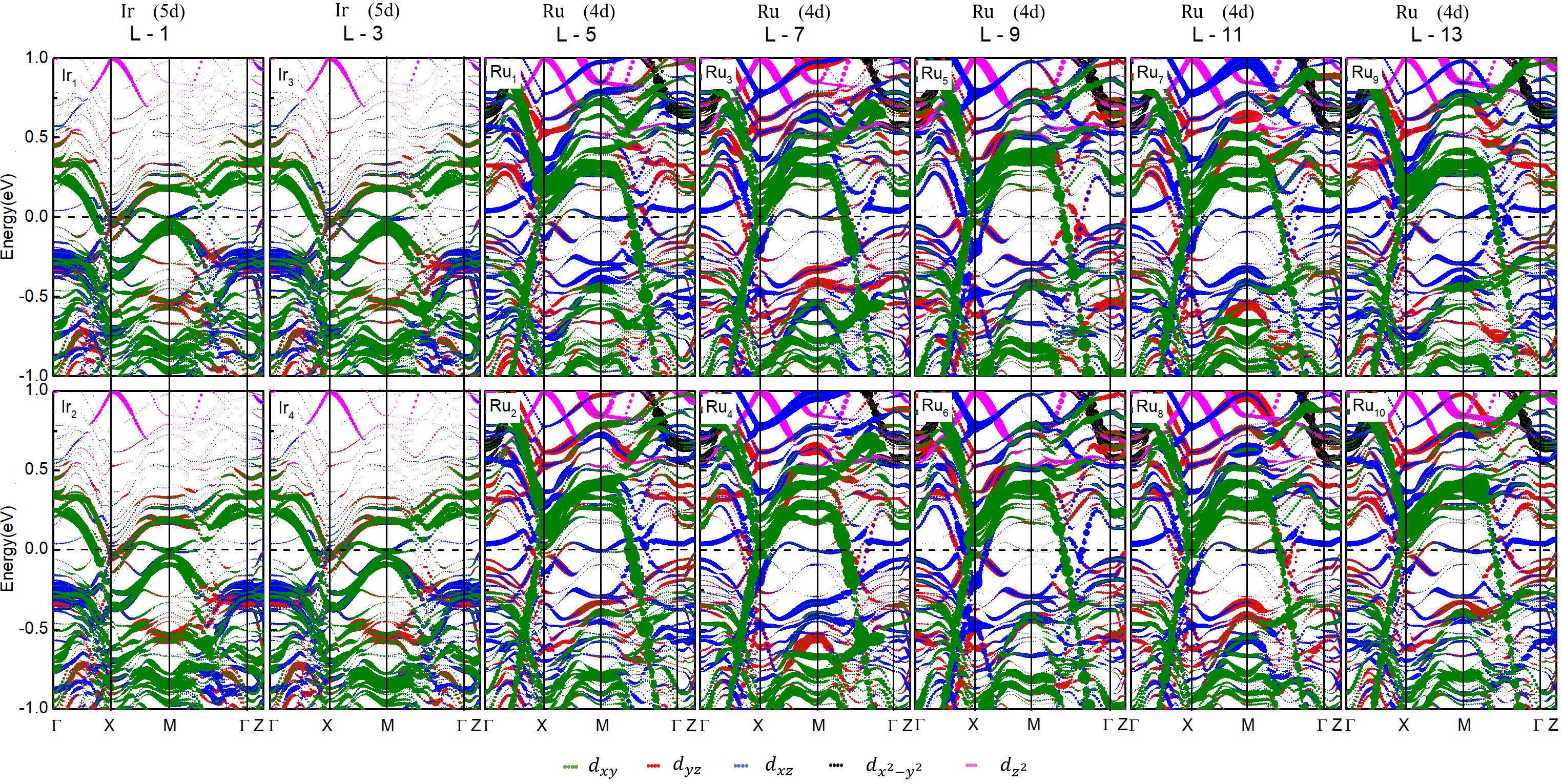}%
	\caption{\label{PBANDS}Band structures  of (SrRuO$_3$)$_5$/(SrIrO$_3$)$_2$(001) showing the contribution of Ir 5$d$ and Ru 4$d$ orbitals in different layers.  The green, red, blue, black and magenta colors denote $d_{xy}$, $d_{yz}$, $d_{xz}$, $d_{x^2-y^2}$, $d_{z^2}$ orbitals, respectively.}
\end{figure*}

\begin{figure}[!htp]
	\includegraphics[width=0.5\textwidth]{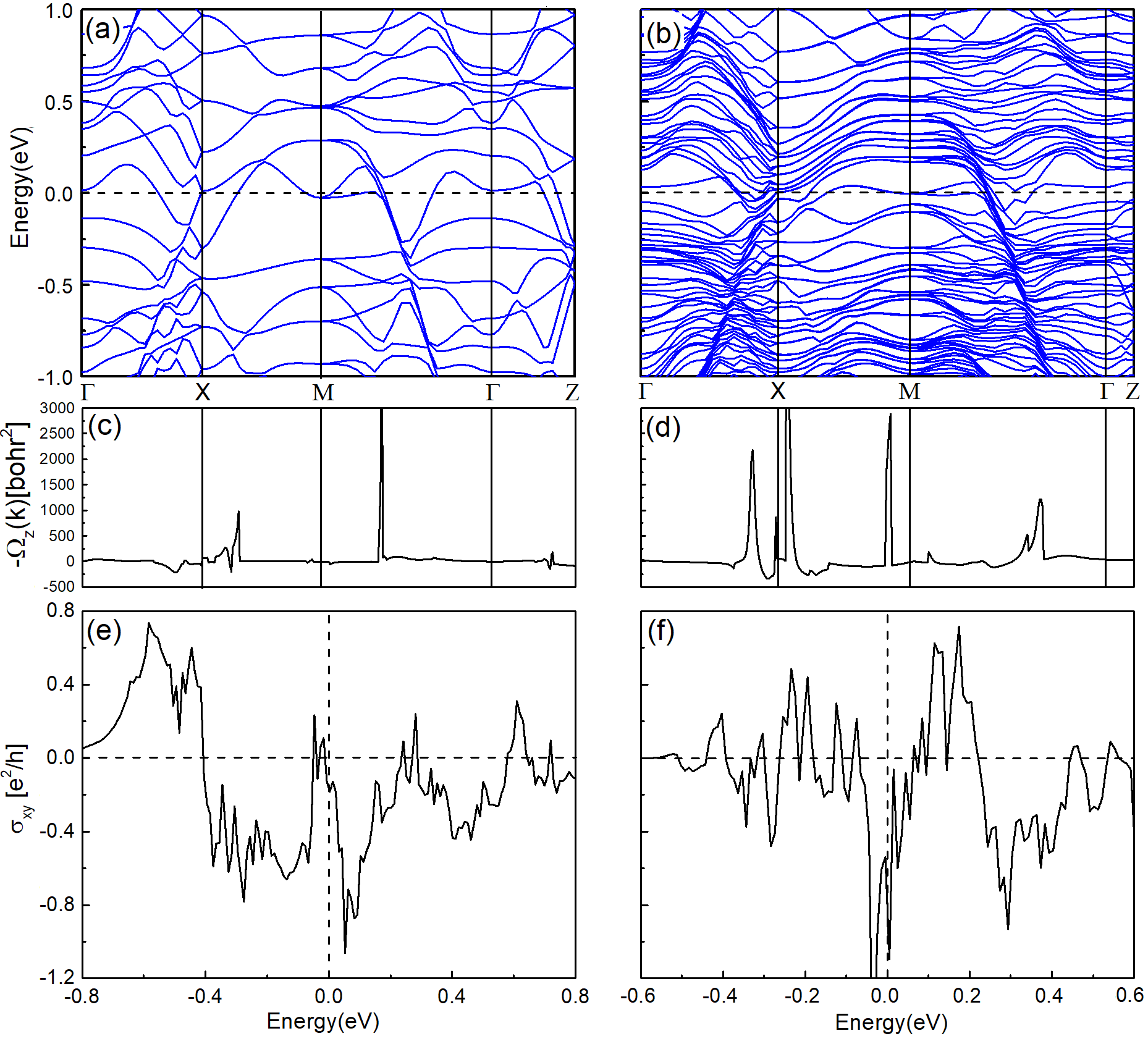}%
	\caption{\label{ahc}Band structures of (a) bulk SrRuO$_3$ and (b) (SrRuO$_3$)$_5$/(SrIrO$_3$)$_2$(001) with SOC with magnetization along the [001]-direction. The Berry curvatures (c-d) are plotted along the same $k$-path, and (e-f) the corresponding anomalous Hall conductivity $\sigma_{xy}$ as a function of the chemical potential in units of $e^2$/h.}
\end{figure}

\begin{figure*}[t!]
	\includegraphics[width=1.0\textwidth]{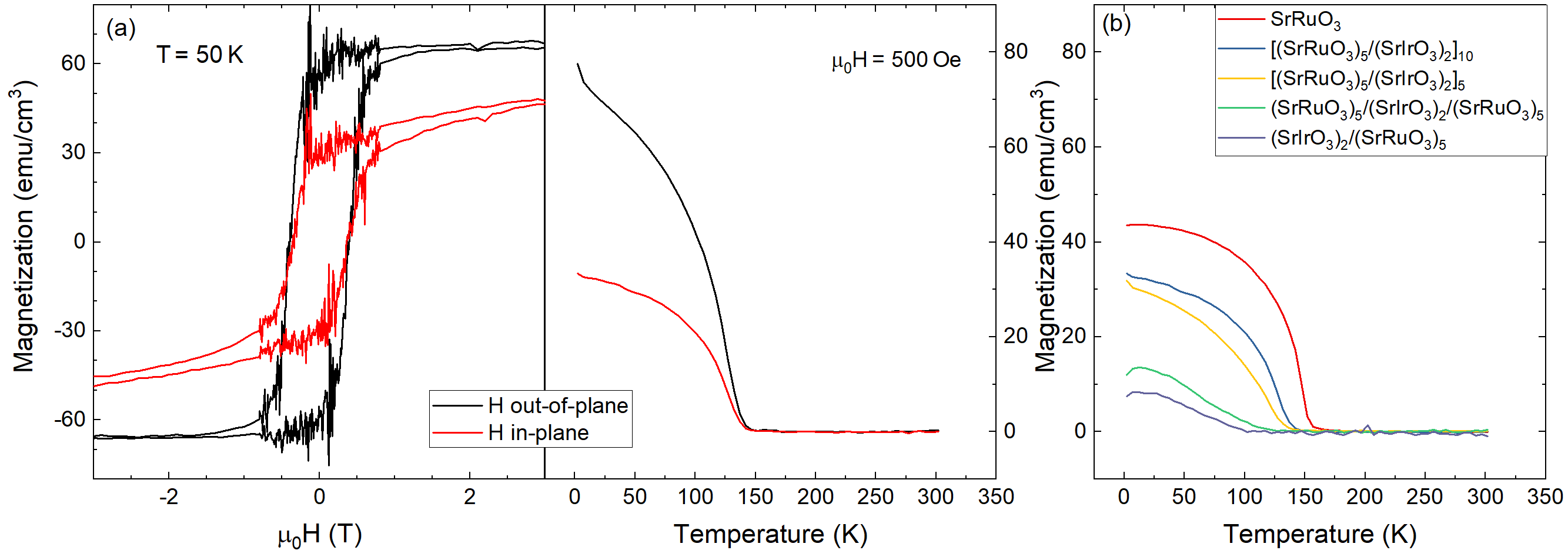}%
	\caption{\label{fig-Magnetization}Temperature- and magnetic field-dependent magnetization of the SROSIO heterostructures. (a) $M(H)$ and $M(T)$ curves for a [(SRO)$_5$/(SIO)$_2$)]$_{10}$ heterostructure with external magnetic field oriented out-of-plane (black) and in-plane (red) with respect to the film surface. The difference in coercitive field indicates a magnetic easy axis orientated out-of-plane. (b) Temperature-dependent magnetization of different [(SRO)$_5$/(SIO)$_2$)]$_{k}$ heterostructures measured in an in-plane magnetic field orientation. As magnetic reference the magnetization of a pure SRO sample is added.}
\end{figure*}

The DFT+$U$+SOC calculations show that bulk SRO is halfmetallic with a spin and orbital magnetic moment of 1.39 \mub~and 0.03 \mub, respectively and SIO is a semimetal with quenched spin and orbital moments, consistent with previous studies \cite{Kee,Franchini}. In Fig.~\ref{DFT_structure} we show  a side view of the spin density (with FM interlayer coupling) and the layer-resolved density of states (LDOS) of the (SrRuO$_3$)$_5$/(SrIrO$_3$)$_2$(001) SL from the DFT+$U$+SOC calculation with magnetization along the $[001]$-direction. The LDOS shows metallic behavior (note that SOC formally entangles the two spin directions, nevertheless predominance of majority spin states at the Fermi level is preserved)  both in the SRO and SIO parts of the SL with a dip of the DOS at the Fermi level. 

Concerning the magnetic properties, the spin and orbital moment at the Ru sites in the inner part of the SRO region are $\sim 1.41$\mub\ and $0.05$\mub, respectively, whereas a slight reduction occurs at the interface, $\sim 1.38$\mub\ and $0.02$\mub. In contrast to bulk SIO, in the SRO/SIO SL a finite spin and orbital moment is induced at the Ir-sites of 0.09 and 0.04\mub, respectively, emphasizing the electronic and magnetic reconstruction in this system. The spin density in Fig.~\ref{DFT_structure} illustrates the \tg\ orbital character at the  Ru-sites, as well as a notable contribution from the oxygen sites indicating substantial covalency of the bond in the SRO part, whereas the spin density is nearly quenched in the SIO part. We have performed calculations for a [(SrRuO$_3$)$_3$/(SrIrO$_3$)$_2$]$_2$(001) SL with ferro- and antiferromagnetic arrangement of the two SRO parts.  Our results indicate that the FM coupling between the SRO blocks through SIO is strongly favored by $\sim 0.652$ eV per simulation cell.

More insight into the electronic properties and origin of the electronic reconstruction in the (SrRuO$_3$)$_5$/(SrIrO$_3$)$_2$(001) SL can be obtained from the band structure plotted in Fig. \ref{PBANDS} which also indicates the layer resolved orbital character. The band structure shows multiple bands crossing the Fermi level, dispersive ones of predominantly \dxy\ character both in the SIO and SRO part and flatter ones of \dxz\ character in the SRO part, in particular in the interface region.

The band structures including SOC with out-of-plane magnetization of the end member SRO and the  (SrRuO$_3$)$_5$/(SrIrO$_3$)$_2$(001) SL are displayed in Fig.~\ref{ahc}(a) and  (b). The band structure of bulk SRO indicates multiple bands crossing at $E_F$, with both electron- and hole-like behavior. Some features of the SRO bulk band structure are visible also in the one of the SL with layer-dependent offsets and superimposed SIO contribution.   By fitting the band structure to a tight binding model with Wannier orbitals using the wannier90 code\cite{Marzari}, we have furthermore calculated the Berry curvature (Fig.~\ref{ahc}c) and (Fig.~\ref{ahc}b) and anomalous Hall conductivity (Fig.~\ref{ahc}e) and (Fig.~\ref{ahc}f) of the two systems. The largest contribution to the Berry curvature $\Omega$($k$) arises between $\Gamma$ and $X$ in bulk SrRuO$_3$ and along $X$-M in (SrRuO$_3$)$_5$/(SrIrO$_3$)$_2$(001). Overall this leads a notable AHC exhibiting sign reversals as a function of the chemical potential. While at the Fermi level the AHC of SrRuO$_3$ is nearly quenched, it is negative for the (SrRuO$_3$)$_5$/(SrIrO$_3$)$_2$(001) SL. As will be shown in section D, this is consistent with the experimental data below 80 K.

\subsection{Magnetization}

To investigate the magnetic anisotropy we applied the external magnetic-field perpendicular (out-of-plane) and parallel (in-plane) to the film surface. Corresponding $M(H)$ and $M(T)$ curves can be found in Fig. \ref{fig-Magnetization}(a). For magnetic fields oriented perpendicular to the film surface the magnetization is 3 times larger than for a field of same magnitude lying in the film plane, indicating an out-of-plane anisotropy of the SROSIO heterostructures. 

\begin{table}[b!]
	\caption{\label{table-Tc-vs-interfaces}Magnetic ordering temperature $T_C$ of [(SRO)$_5$/(SIO)$_2$)]$_k$ heterostructures with different number of interfaces in comparison with that of a pure 42 monolayer thick SrRuO$_3$ thin film.}
	\begin{ruledtabular}
		\begin{tabular}{ccc}
			Sample & Interfaces & $T_C$ (K)\\
			\hline
			SrRuO$_3$ & - & 157\\
			$[$(SrRuO$_3$)$_5$/(SrIrO$_3$)$_2]_{10}$ & 19 & 145\\
			$[$(SrRuO$_3$)$_5$/(SrIrO$_3$)$_2]_{5}$ & 9 & 137\\
			(SrRuO$_3$)$_5$/(SrIrO$_3$)$_2$)/(SrRuO$_3$)$_5$ & 2 & 126\\
			(SrIrO$_3$)$_2$/(SrRuO$_3$)$_5$ & 1 & 102\\
		\end{tabular}
	\end{ruledtabular}
\end{table}

To investigate the influence of the number of interfaces we measured temperature-dependent magnetization curves for [(SRO)$_5$/(SIO)$_2$)]$_k$ thin films with different repetitions $k$ of the bilayer structure in an in-plane oriented magnetic field of $\mu_0H=500$\,Oe. As magnetic reference signal we use a pure SRO thin film. In Fig \ref{fig-Magnetization}(b) the obtained $M(T)$-curves are shown. Compared to the SRO reference all SROSIO heterostructures have a smaller magnetization and a lower ordering temperature $T_C$.

This can be explained by taking into account the ordering of the spins close to the interface. Due to the broken inversion symmetry between SRO and SIO a DM interaction results which forces the spins in a non-collinear structure \cite{Soumyanarayanan2016, Cho2015}. For most of the SRO layers, which are sandwiched symmetrically by SIO layers on top and bottom, the DM interaction has the same strength but opposite sign at the top and bottom. Thus, for magnetic fields perpendicular to the interface their effect cancels out and the magnetic moments of those SRO layers align ferromagnetically like in a pure SRO thin film. However, for SRO layers with just one interface to SIO, such as the first layer grown directly on the STO substrate and the last top SRO layer for the sample with only 1 SIO layer between 2 SRO layers, the DMI does not cancel and the spins align in a non-collinear spin structure. For SROSIO heterostructures this means, that in heterostructures with $k>1$ repetitions ($k-1$) SRO layers order like a pure SRO thin film. With increasing $k$ values the ordering temperature $T_C$ of the heterostructure therefore approaches that of the pure SRO (see Tab. \ref{table-Tc-vs-interfaces}). 

In an in-plane oriented magnetic field the spins order at the top and bottom interface in a conical phase with opposite rotation sense. Due to the rotation of the moments the magnetization parallel to the film plane is reduced compared to a pure SRO film. For example the bilayer with 5 monolayers SRO has the same $T_C$ as 4 monolayers of a pure SRO thin film \cite{Ishigami2015}.

\subsection{Magnetotransport}
\subsubsection{\label{sec-temperature-dependent-Hall-effect}Temperature dependent Hall effect and magnetoresistance}

\begin{figure}[b!]
	\includegraphics[width=0.5\textwidth]{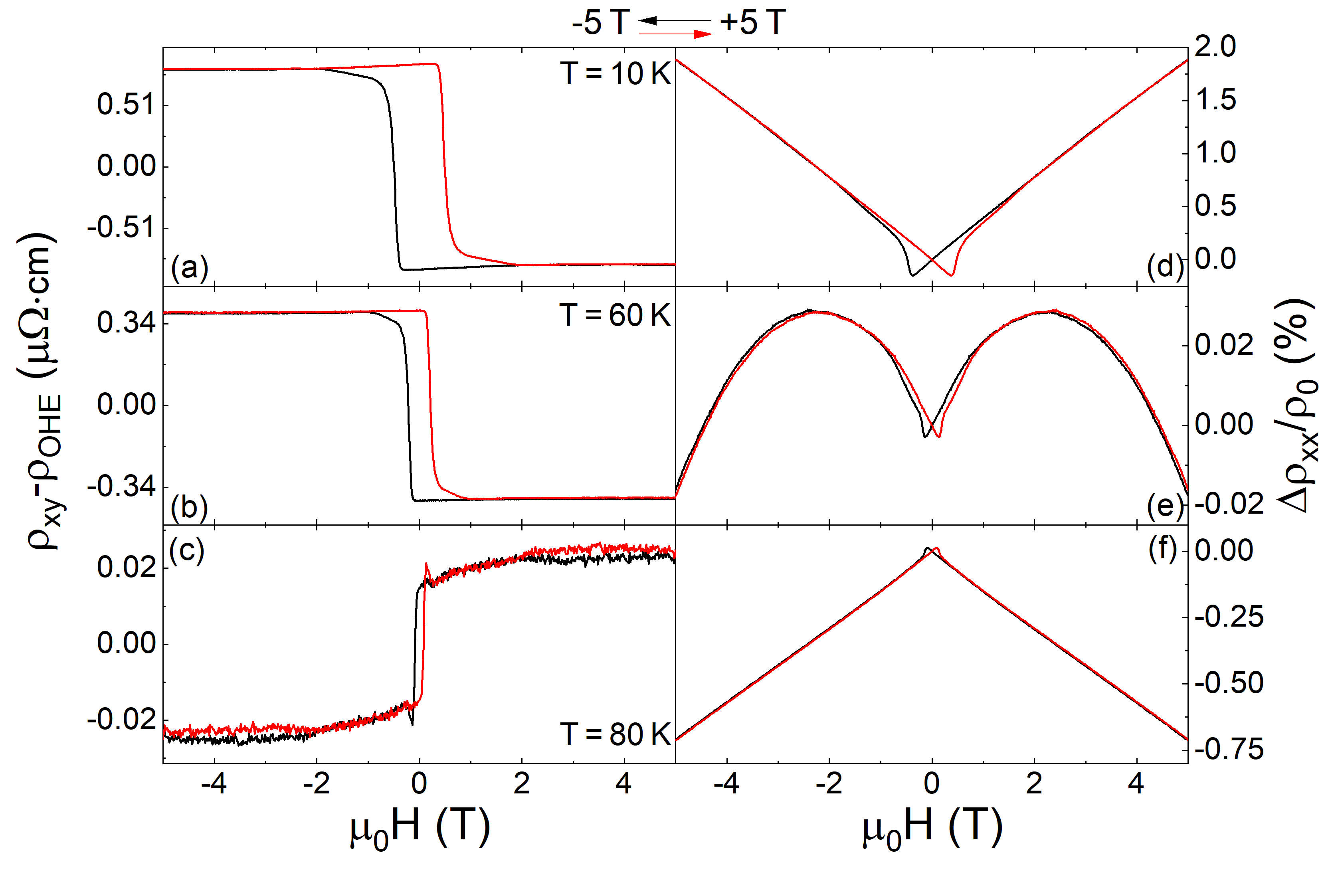}%
	\caption{\label{fig-T-dependence-of-Hall-and-MR}Hall effect (a)-(c) and isothermal magnetoresistance (d)-(f) at various temperatures for a [(SRO)$_5$/(SIO)$_2$)]$_{10}$ heterostructure measured in out-of-plane oriented external magnetic field. Black lines indicate magnetic field sweeps from $+5$\,T to $-5$\,T and red lines the opposite sweep direction. For the Hall resistance the contribution of the ordinary Hall effect was subtracted.}
\end{figure}

To investigate the magnetotransport properties of our heterostructures we applied an external magnetic field in out-of-plane direction and measured both Hall resistance $\rho_{xy}$ (transversal to current $j$) and isothermal magnetoresistance MR $=(\rho_{xx}(H)-\rho_{xx}(0))/\rho_{xx}(0)$ (parallel to $j$). The contribution of the ordinary Hall effect (OHE) was determined by a linear fit in the high-field regimes ($|\mu_0H|>4$\,T) and subtracted from the total Hall resistance. As shown in Fig. \ref{fig-T-dependence-of-Hall-and-MR} (a)-(c) the anomalous Hall effect (AHE) changes its sign between $T=$ 60\,K and $T=$ 80\,K. This sign change is known for SrRuO$_3$ and was related to a band-structure effect \cite{Haham2011}.

\begin{figure}[t!]
	\includegraphics[width=0.46\textwidth]{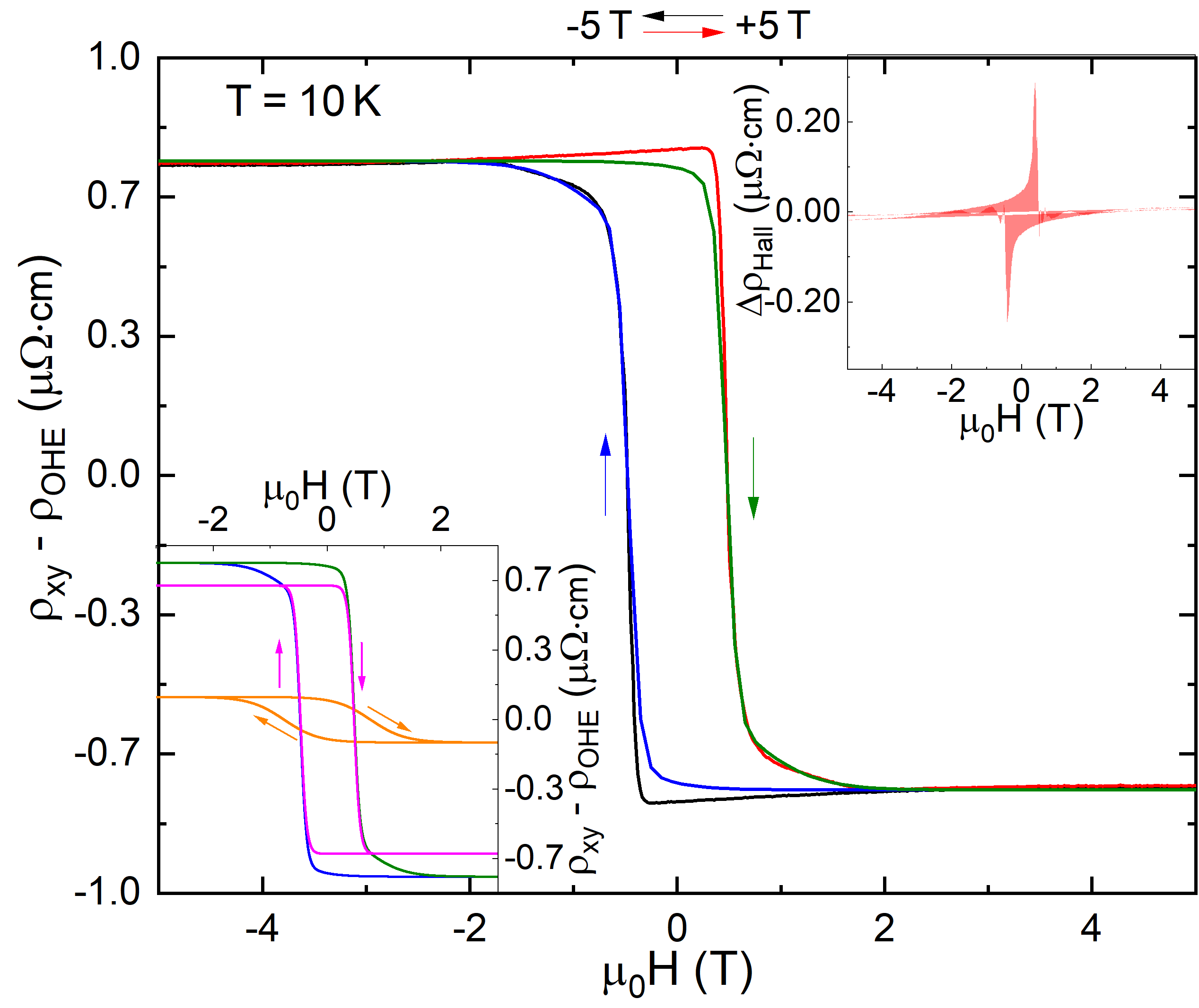}%
	\caption{\label{fig-80K-Hall-2-Channel-Effect}Hall effect of a [(SRO)$_5$/(SIO)$_2$)]$_{10}$ heterostructure (black and red lines) and the adjusted two channel AHE model (blue and green) at $T=$ 10\,K. The contributions of both channels are visualized in the lower inset. The upper inset shows the difference $\Delta\rho_{\text{Hall}}=\rho_{\text{Data}}-\rho_{\text{Fit}}$ between experiment and fit.}
\end{figure}

The shape of the AHE can be described by a phenomenological model with two AHE contributions,
\begin{displaymath}
\rho_{AHE}\propto \rho_{s1}\cdot\tanh\left(\frac{H-H_{c1}}{H_{s1}}\right)+\rho_{s2}\cdot\tanh\left(\frac{H-H_{c2}}{H_{s2}}\right),
\end{displaymath}

introduced by \cite{Groenendijk2018, Thiel2020}. In Fig. \ref{fig-80K-Hall-2-Channel-Effect} this two-channel model is fitted to the Hall effect at $T =$ 10\,K. Here, both AHE contributions have similar sign but different sizes and coercitive fields (cf. the lower inset). 

The upper inset shows the difference $\Delta\rho_{\text{Hall}}=\rho_{\text{Data}}-\rho_{\text{Fit}}$ between experimental data and the fit with the two channel model. A maximal value of $\Delta\rho_{\text{Hall}} =\pm 0.29$\,$\upmu\Omega\text{cm}$ is found at $\mu_0H\sim\pm0.121$\,T. This additional contribution $\Delta\rho_{\text{Hall}}$ to the total Hall effect may hint at a topological Hall effect (THE) in our heterostructures (see below). The temperature-dependence of the coercitive fields $H_{c1}$ and $H_{c2}$ obtained from the fits by the two channel AHE function is listed in the Supplemental Material \cite{Supplemental}. 

\begin{figure}[b!]
	\includegraphics[width=0.5\textwidth]{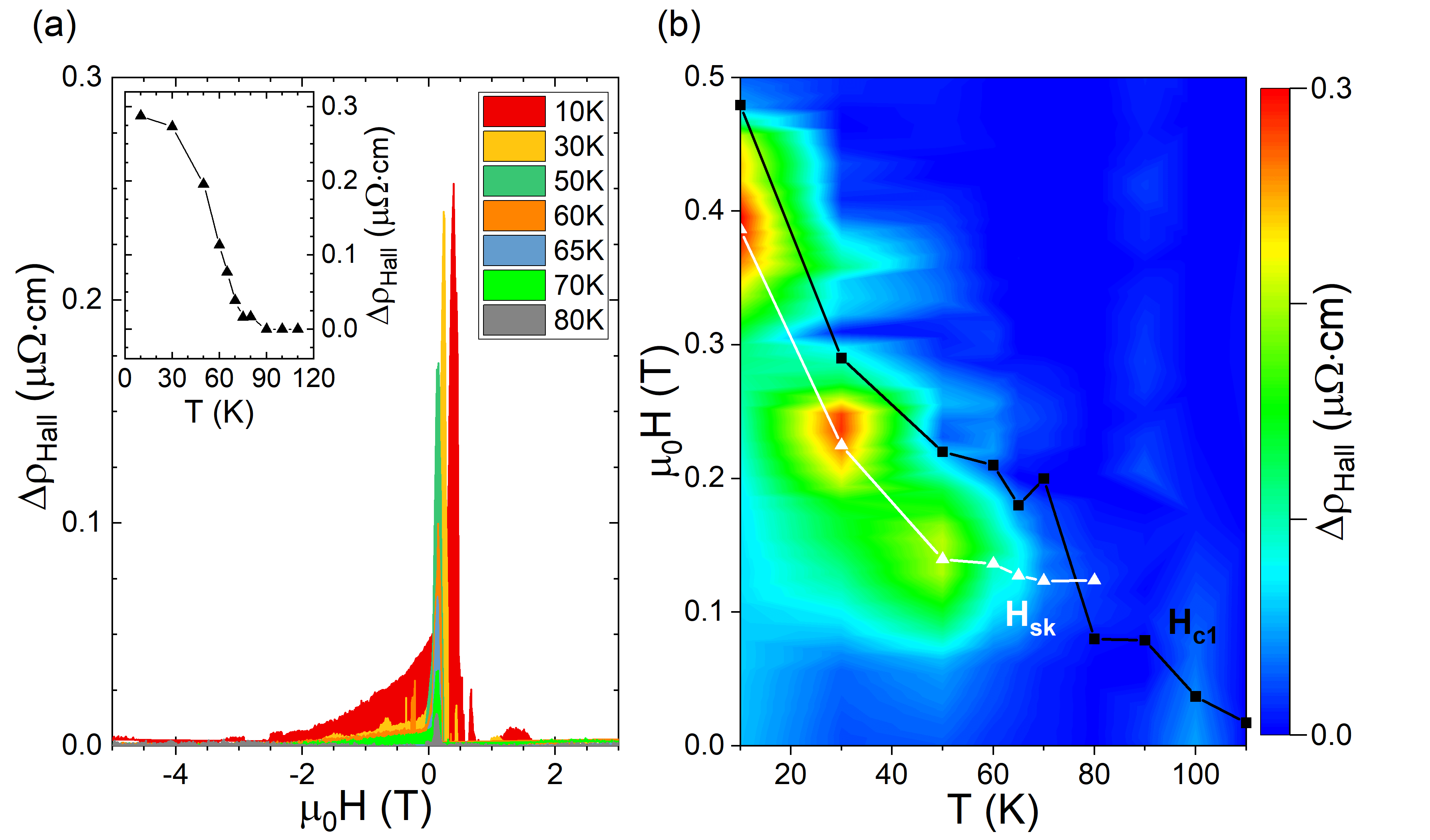}%
	\caption{\label{fig-T-dependence-deltaRho}Magnetic field dependence of $\Delta\rho_{\text{Hall}}$ (see text) of a [(SRO)$_5$/(SIO)$_2$)]$_{10}$ heterostructure for temperatures $10\,\text{K}\leq T\leq80\,\text{K}$. The inset shows the temperature dependence of the maximal absolute value of $\Delta\rho_{\text{Hall}}$. (b) Color map of $\Delta\rho_{\text{Hall}}$ in the $T-H$ plane. Black squares and white triangles  denotes the coercitive field ($H_{c1}$) and the field ($H_{sk}$) at which the additional contribution $\Delta\rho_{\text{Hall}}$ reaches its maximum, respectively.}
\end{figure}

For different temperatures the magnetic field dependence of the absolute value of this extra contribution $\Delta\rho_{\text{Hall}}$ is shown in Fig. \ref{fig-T-dependence-deltaRho}. We observe a shift of the maximum from $\mu_0H\sim 0.4$\,T at $T=$ 10\,K to $\mu_0H\sim 0.12$\,T at $T=$ 80\,K and a decrease from $\Delta\rho_{\text{Hall}} = 0.29$\,$\upmu\Omega\text{cm}$ to $\Delta\rho_{\text{Hall}} = 0.016$\,$\upmu\Omega\text{cm}$. For temperatures above $T\geq90$\,K $\Delta\rho_{\text{Hall}}$ vanishes.

We denote the field at which the absolute value of the additional Hall contribution is maximal as $H_{sk}$. The latter is located close to the coercitive field $H_{c1}$ at all temperatures below 90~K. The temperature dependence of the maximal absolute size of $\Delta\rho_{\text{Hall}}$ is shown in the inset of Fig. \ref{fig-T-dependence-deltaRho}.



%

Before analyzing the Hall effect and its angular dependence further, we first turn to the magnetoresistance (MR). The temperature dependent MR (Fig. \ref{fig-T-dependence-of-Hall-and-MR} (d)-(f) can be separated in two ranges. At low temperatures ($T=$ 10\,K) the low field MR is negative and becomes positive at higher magnetic fields whereas it looks opposite at higher temperatures ($T=$ 80\,K). The crossover between the two types of dependences is located at $T=$ 60\,K.


\begin{figure}[t!]
	\includegraphics[width=0.5\textwidth]{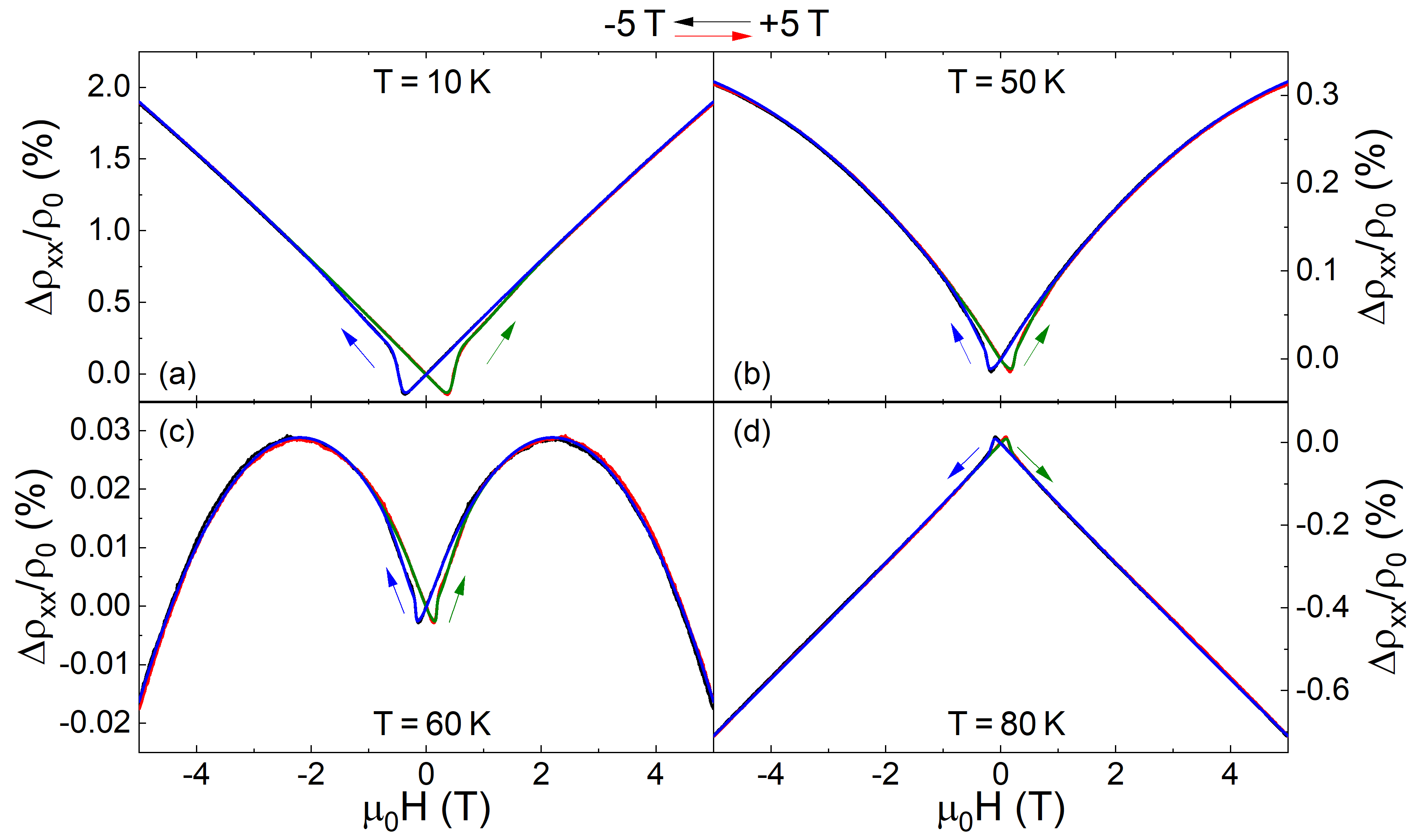}%
	\caption{\label{fig-T-dependence-MR-with-Fit}Out of plane MR (black and red) and two channel MR-model (blue and green) of a [(SRO)$_5$/(SIO)$_2$)]$_{10}$ heterostructure for $T=$ 10\,K (a), 50\,K (b), 60\,K (c) and 80\,K (d).}
\end{figure}

For modeling the MR we can also use a two channel model

\begin{eqnarray}
\frac{\Delta\rho_{xx}}{\rho_{0}}&\propto& A\Bigl[H+M_{s1}\cdot\tanh\left(\frac{H-H_{c1}}{H_{s1}}\right)\Bigr]^2\nonumber\\ &+& C\Bigl[H+M_{s2}\cdot\tanh\left(\frac{H-H_{c2}}{H_{s2}}\right)\Bigr]^2
\end{eqnarray}

with two additive contributions \cite{Supplemental}. As shown in Fig. \ref{fig-T-dependence-MR-with-Fit} the experimentally obtained MR can be well described by this two channel model, where the two channels have the same coercitive fields $H_{ci}$ as used for the fits of the anomalous Hall effect.

\subsubsection{Field-angle dependent Hall effect and magnetoresistance}

Study of the angular dependence of the Hall effect, in particular its additional contribution $\Delta\rho_{\text{Hall}}$ discussed above, enables to investigate the stability of the assumed skyrmion structure. In our experiments the magnetic-field was tilted in two different perpendicular arrangements from out-of-plane towards in-plane direction. When the in-plane component ($H_{\text{in-plane}}$) of the magnetic-field is tilted parallel to the current $j$ the angle is labeled $\theta$, whereas we used $\beta$ as tilt angle for the direction with both $H_{\text{in-plane}}$ and $H_{\text{out-of-plane}}$ perpendicular to $j$ (cf. the sketch in Fig. S6 of Supplemental Material \cite{Supplemental}). For the out-of-plane direction, as described in Sec. \ref{sec-temperature-dependent-Hall-effect}, $\theta = 0^\circ$ and $\beta = 0^\circ$. Comparing both directions one finds similar shapes of the Hall effect except for angles near $90^\circ $ where the external magnetic-field lies in the film plane. In particular the ordinary Hall constant $R_0$ is similar for both tilt angles (see Supplemental Material \cite{Supplemental}), indicating that it is determined by the normal field component.

\begin{figure}[t!]
	\includegraphics[width=0.5\textwidth]{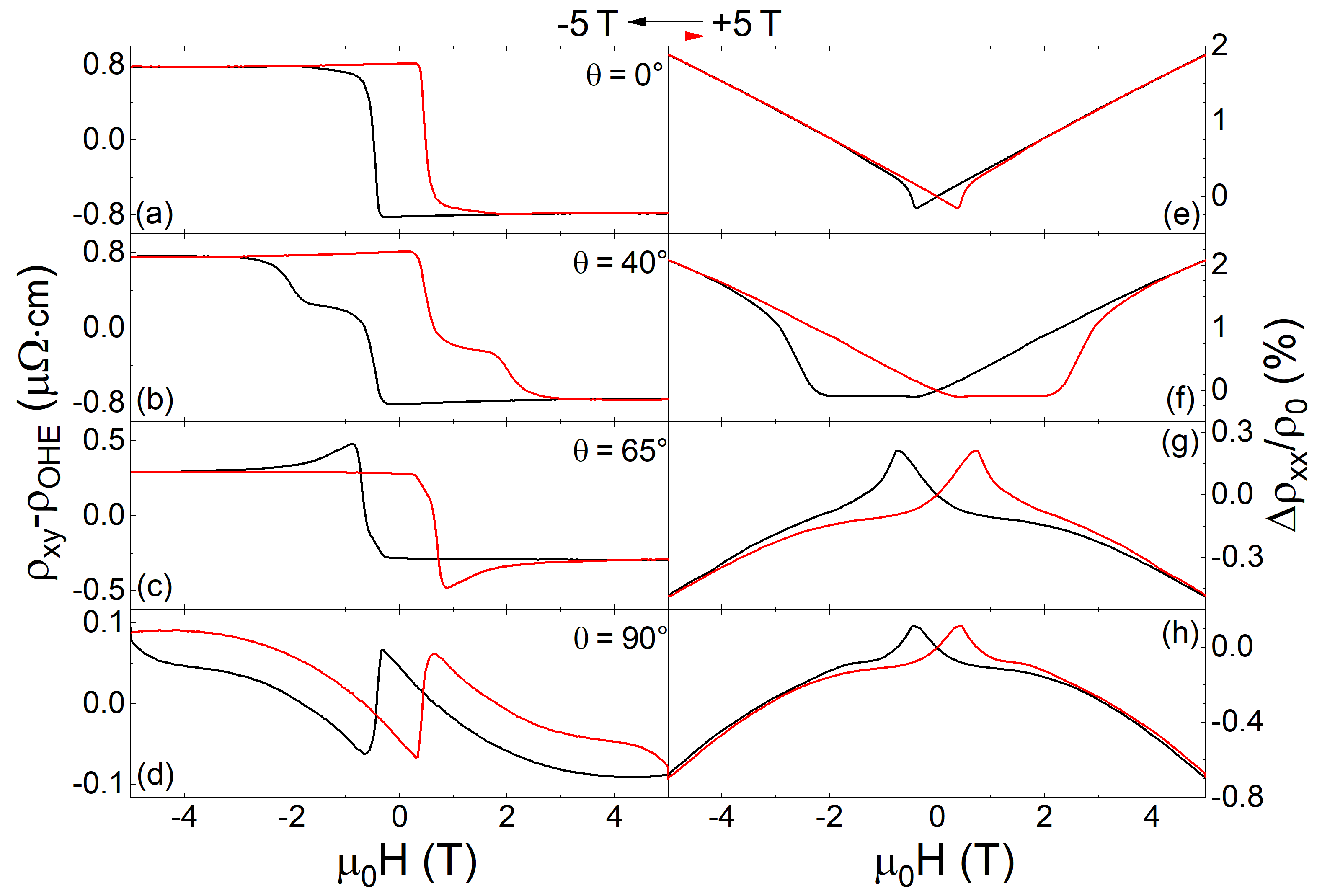}%
	\caption{\label{fig-theta-dependence-Hall-effect-and-MR-10K}Hall effect (a)-(d) and MR (e)-(h) of a [(SRO)$_5$/(SIO)$_2$)]$_{10}$ heterostructure for various field angles $\theta$ from the sample normal to the in-plane direction parallel to the current at $T=$ 10\,K.}
\end{figure}

\begin{figure}[b!]
	\includegraphics[width=0.5\textwidth]{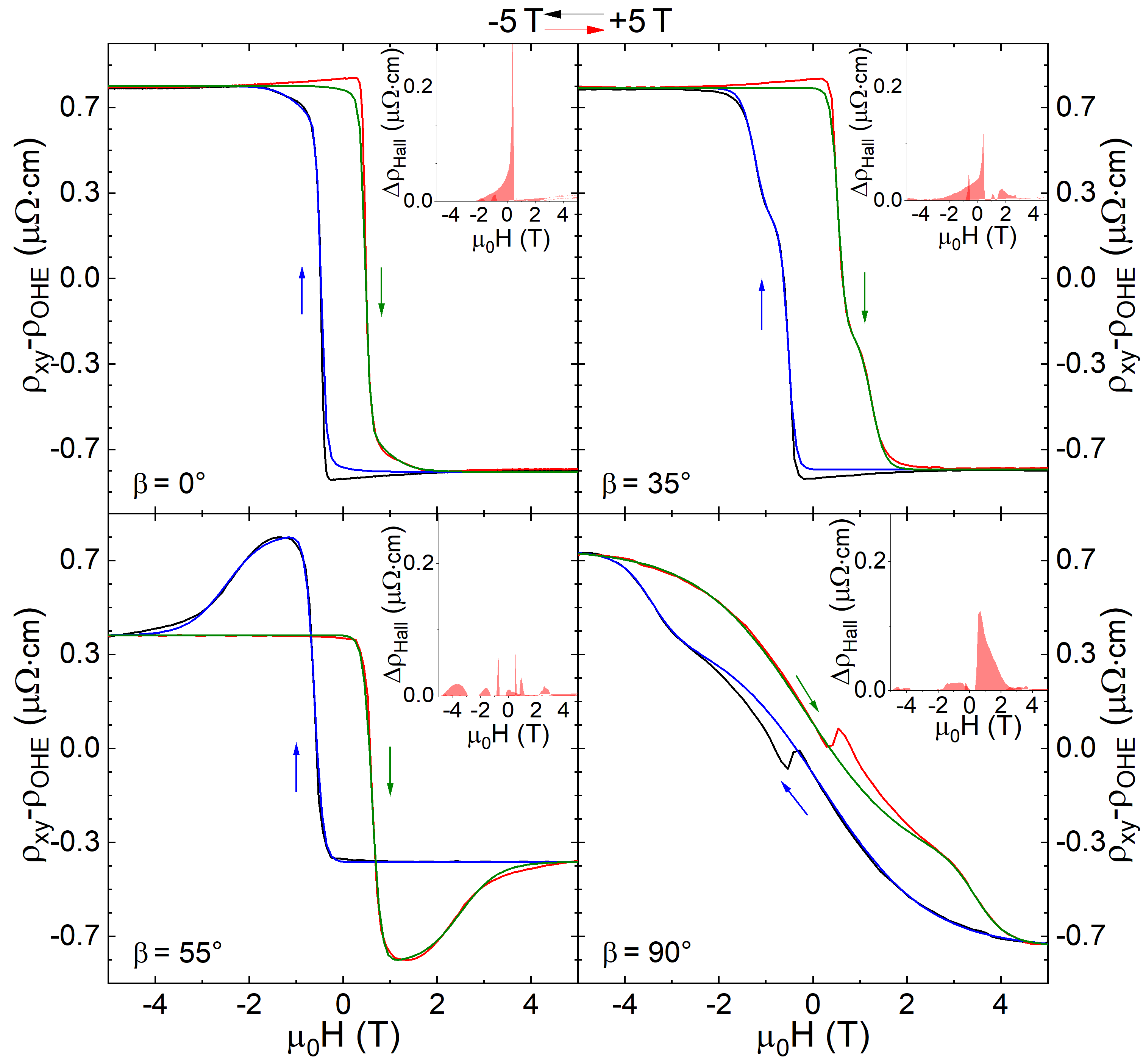}%
	\caption{\label{fig-2-Channel-model-for-beta-dependent-Hall-effect}Hall effect of a [(SRO)$_5$/(SIO)$_2$)]$_{10}$ heterostructure after subtraction of normal contribution for various field angles $\beta$ from the sample normal to the in-plane direction perpendicular to the current (black and red) as well as fits by two-channel model (see text), indicated by blue and green lines. The inset shows the difference $\Delta\rho_{\text{Hall}}$ between data and fit (see text).}
\end{figure}

\begin{figure}[t!]
	\includegraphics[width=0.48\textwidth]{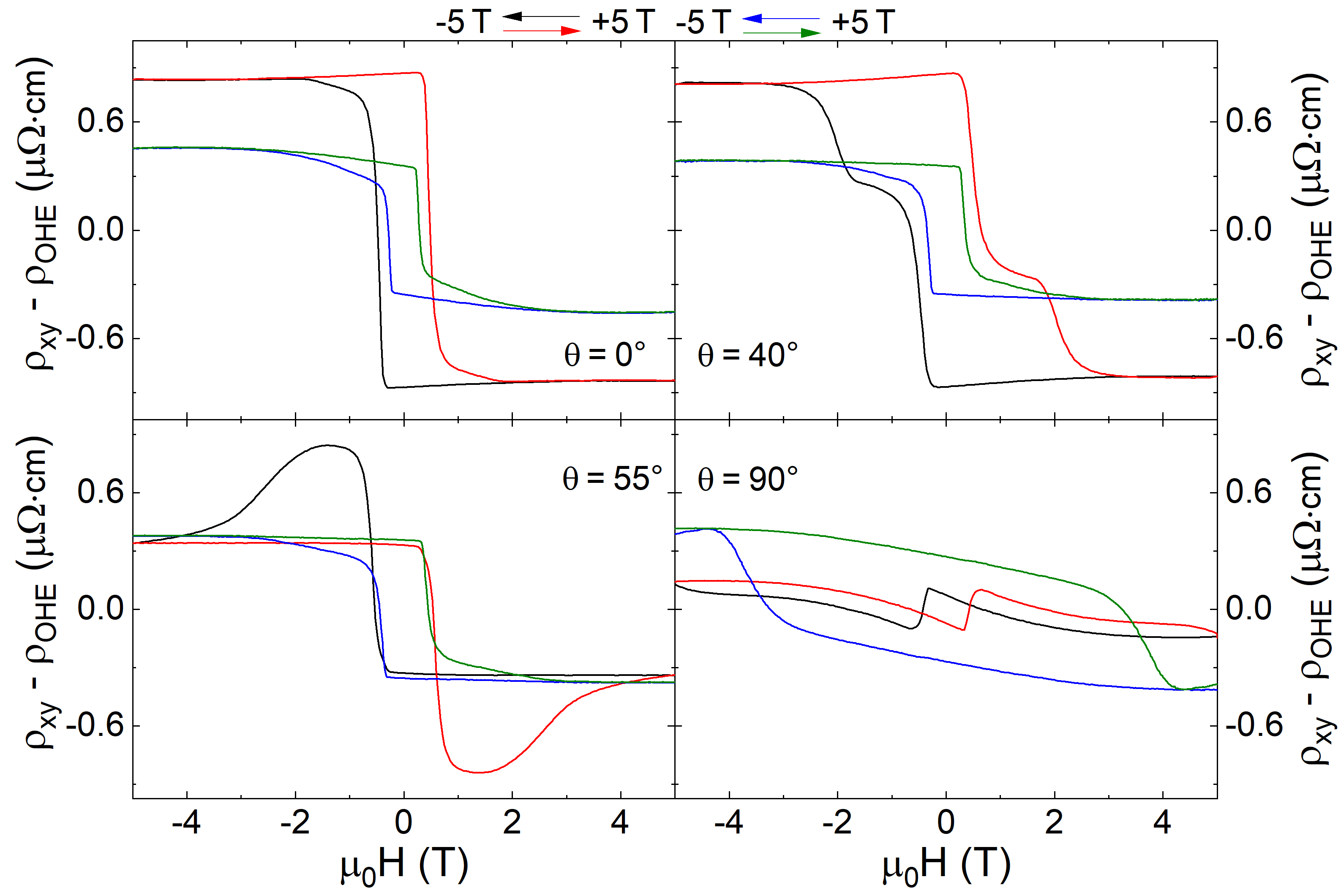}%
	\caption{\label{fig-theta-dependence-Hall-effect-10K-pureSRO-SIOSRO10}Comparison of angle-dependent Hall effect between [(SRO)$_5$/(SIO)$_2$)]$_{10}$ (black and red curves) and pure SRO thin film (blue and green curves) for different tilt angles $\theta$.}
\end{figure}

\begin{figure}[b!]
	\includegraphics[width=0.5\textwidth]{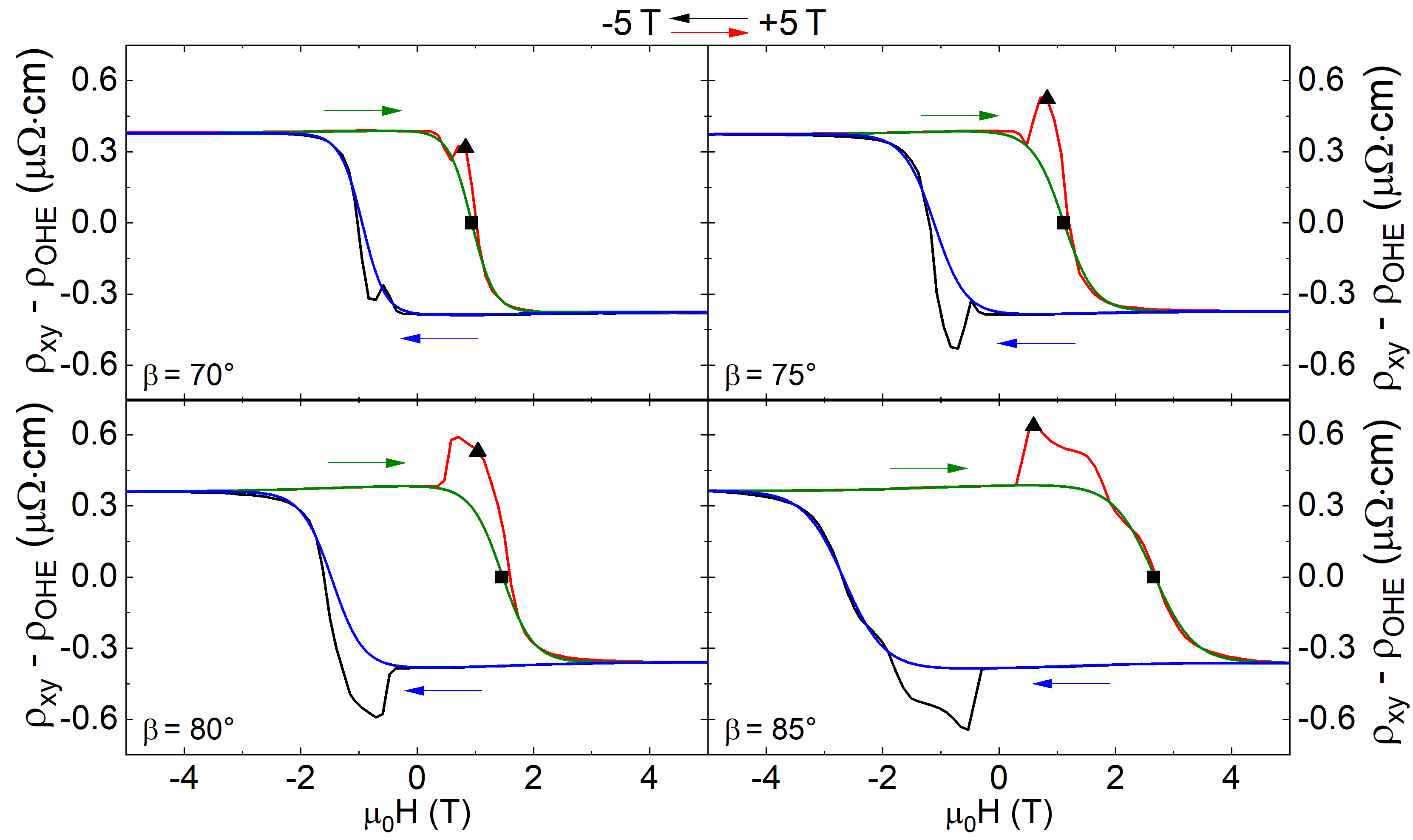}%
	\caption{\label{fig-10K-70deg-85deg-beta-Hall}Hall effect of a [(SRO)$_5$/(SIO)$_2$)]$_{10}$ heterostructure (black and red lines) and the adjusted two channel AHE model (blue and green) at $T=$ 10\,K for field orientations close to an in-plane direction. Note the additional peak-like structure, which cannot be modeled by the two channel AHE model. Squares and triangles denote values of the coercitive field ($H_{c1}$) and the field ($H_{sk}$) at which the additional contribution $\Delta\rho_{\text{Hall}}$ reaches its maximum, respectively.}
\end{figure}

\begin{figure*}[t!]
	\includegraphics[width=1.0\textwidth]{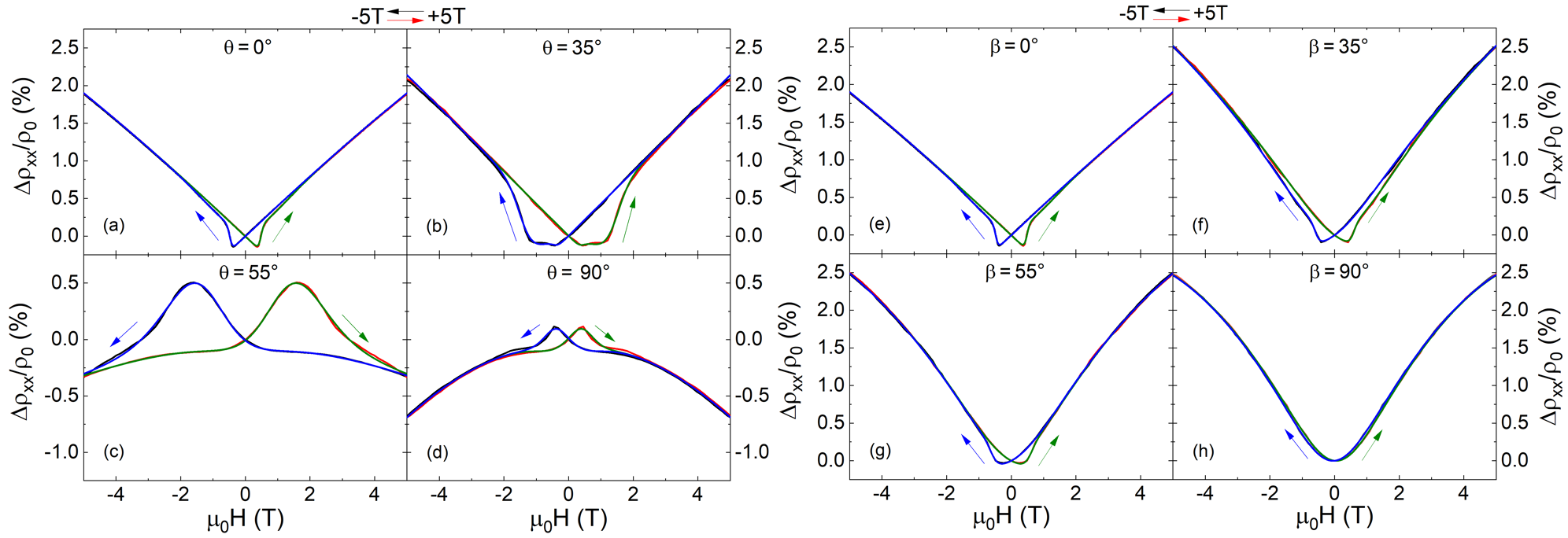}%
	\caption{\label{fig-theta-beta-angle-dependence-MR-with-Fit}Angle-dependence of magnetoresistance (black and red) and two-channel fit (blue and green) at 10\,K for various tilt angles parallel to the current direction: (a) $\theta=0^\circ$, (b) $\theta=35^\circ$, (c) $\theta=55^\circ$, (d) $\theta=90^\circ$, and perpendicular to the current direction: (e) $\beta=0^\circ$, (f) $\beta=35^\circ$, (g) $\beta=55^\circ$ and (h) $\beta=90^\circ$.}
\end{figure*}

Similar as above, we fit the AHE at various tilt angles $\beta$ by the two channel anomalous Hall effect model, cf. Fig. \ref{fig-2-Channel-model-for-beta-dependent-Hall-effect}. For low angles up to $\beta=35^\circ$ both channels contribute with similar sign, while for $\beta=55^\circ$ they have opposite sign, producing a strong hump-like feature in $\rho_{\text{AHE}}$. The insets of Fig. \ref{fig-2-Channel-model-for-beta-dependent-Hall-effect} display the additional contribution $\Delta\rho_{\text{Hall}}$, whose field dependence, as will be discussed below, can be used to analyze the stability of possible interface skyrmions. 

We note, that the Hall effect of a pure SRO thin film demonstrates a weaker dependence of the field-angle compared to the (SROSIO)$_{10}$ heterostructure (Fig. \ref{fig-theta-dependence-Hall-effect-10K-pureSRO-SIOSRO10}). Furthermore, the Hall effect of SRO can be consistently described by the two channel anomalous Hall effect model (cf. Kim et al. \cite{Kim2020} and Supplemental Material \cite{Supplemental}) without any additional contribution. Thus, $\Delta\rho_{\text{Hall}}$ must be related to the interface between SRO and SIO.

\begin{figure}[b!]
	\includegraphics[width=0.5\textwidth]{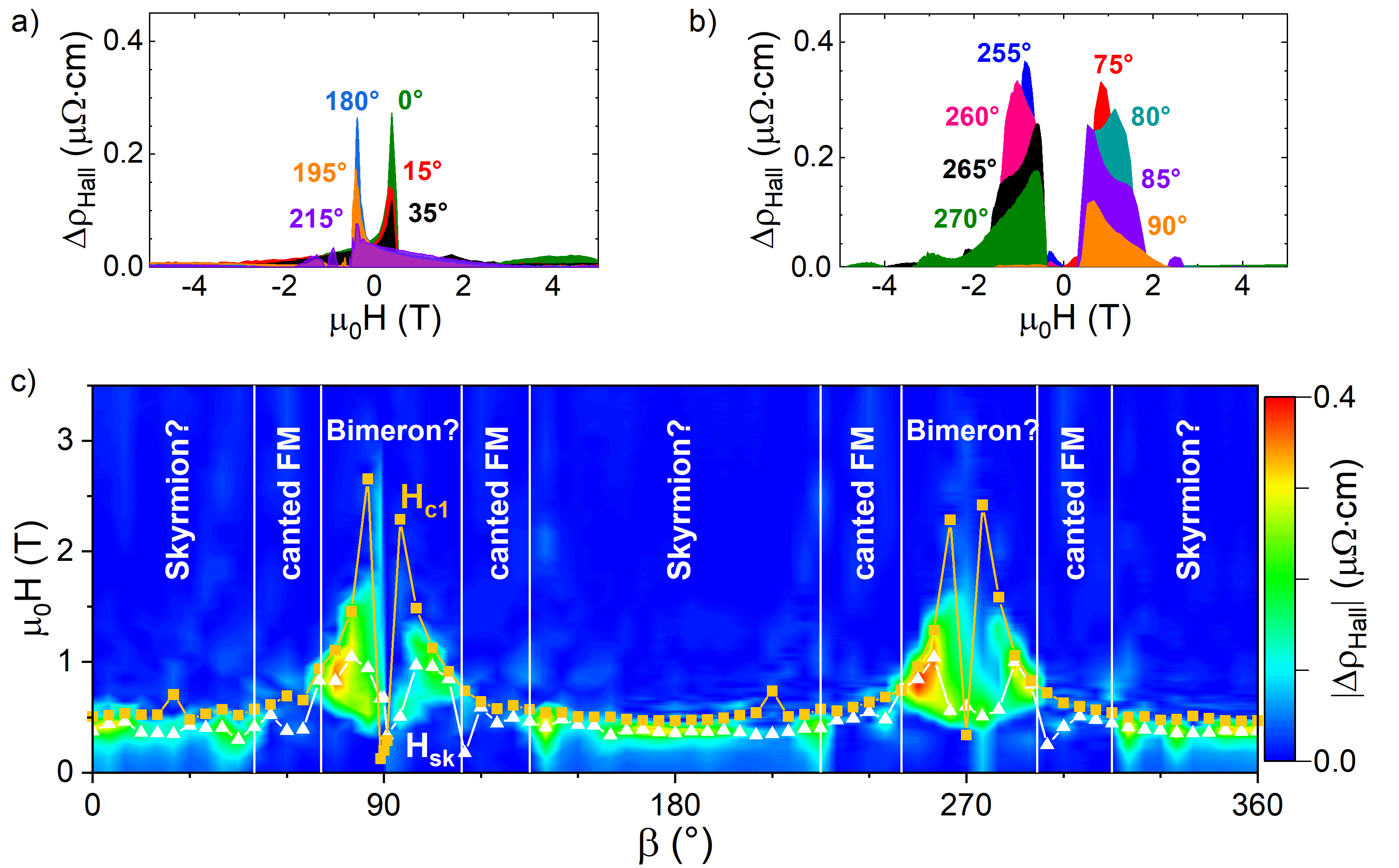}%
	\caption{\label{fig-180-deg-symmetry-deltaRho-and-beta-dependent-deltaRho}Additional contribution $\Delta\rho_{\text{Hall}}$ (cf. Fig.~\ref{fig-10K-70deg-85deg-beta-Hall}) at 10~K for different magnetic field angles $\beta$ as indicated by the labels (a) and (b), close to the film normal and in-plane field orientation, respectively. (c)  Color map of $\Delta\rho_{\text{Hall}}$ in the $\beta-H$ plane, which include possible N{\'{e}}el skyrmions (Skyrmion), canted ferromagnetic phase (canted FM) and possible Bimerons  } 
\end{figure}

Comparing $\Delta\rho_{\text{Hall}}$ for all measured angles $\beta$ we observe the expected "$180^\circ$ symmetry", as demonstrated in Fig. \ref{fig-180-deg-symmetry-deltaRho-and-beta-dependent-deltaRho}(a)-(b).

Assuming a topological Hall effect due to skyrmions as origin of $\Delta\rho_{\text{Hall}}$, we now analyze its angular dependence. For Bloch-type skyrmions in MnSi thin films, Yokouchi {\it et al.} \cite{Yokouchi2014} found a stable skyrmion region for film thickness $d$ smaller than the helix length $\lambda$. The expected behavior of interface N{\'{e}}el-type skyrmions in a tilted magnetic field is different. While Bloch skyrmions co-align their axis to the direction of the tilted magnetic field,  N{\'{e}}el skyrmions lock their symmetry axis to the high symmetry direction of the thin film (perpendicular to the interface) leading to a deformation in the direction of the in-plane component ($H_{\text{in-plane}}$) of the magnetic field and thus an increase of their size \cite{Leonov2017}.


Since the magnitude of the THE is inversely proportional to the skyrmion cross section~\cite{Nagaosa2013,Ohuchi2018}, an increasing skyrmion size is expected to lead to a decreasing THE. Importantly, we oberve for increasing field angles within $45^\circ$ from the sample normal (cf. Fig. \ref{fig-180-deg-symmetry-deltaRho-and-beta-dependent-deltaRho}(a)) that the additional Hall contribution indeed shows a decreasing peak size.
This is a remarkable result and supports the interpretation of the $\Delta\rho_{\text{Hall}}$ contribution as THE arising from skyrmions. Outside these regimes, the strong increase of the calculated $r_{sk}$ indicates that interface skyrmions become unstable. In a range of $\pm 20^\circ$ around the in-plane field orientation ($\beta=90^\circ$ and $\beta=270^\circ$) an unexpected peak-like structure appears in the Hall effect close to the coercitive field $H_{c1}$ (Fig. \ref{fig-10K-70deg-85deg-beta-Hall}). This structure could be a hint for a non-trivial spin structure different to N{\'{e}}el-type skyrmions (e.g. magnetic bimerons \cite{Moon2019, Goebel2019} or magnetic bubbles \cite{Zhang2020}).

As shown in Fig. \textcolor{red}{\ref{fig-theta-beta-angle-dependence-MR-with-Fit}}, the MR at 10 K for tilting the magnetic field parallel to the current (tilt angle $\theta$, (a)-(d)), as well as perpendicular to the current (tilt angle $\beta$, (e)-(h)) can always be well described by the two channel MR model. In the former case, a crossover from positive MR for small $\theta$ to negative MR at large fields for large $\theta$ can be observed (cf. also Fig. \ref{fig-theta-dependence-Hall-effect-and-MR-10K} (e)-(h)). The negative longitudinal magnetoresistance (LMR) could be a hint for the manifestation of Weyl fermions due to a chiral imbalance, as observed by Takiguchi \textit{et al.} \cite{Takiguchi2020} in high-quality SRO thin films. Microscopic disorder as for example interface roughness could also lead to a negative LMR \cite{Xu2019}. When tilting the magnetic field perpendicular to current (tilt angle $\beta$) the MR shows a completely different behavior (see Fig. \ref{fig-theta-beta-angle-dependence-MR-with-Fit}(e)-(h)). There appears no crossover from positive to negative MR. Instead the positive MR value at $\pm$5\,T increases and the hysteresis at smaller field values decreases from out-of-plane to in-plane direction.

\begin{figure*}[t!]
	\includegraphics[width=1.0\textwidth]{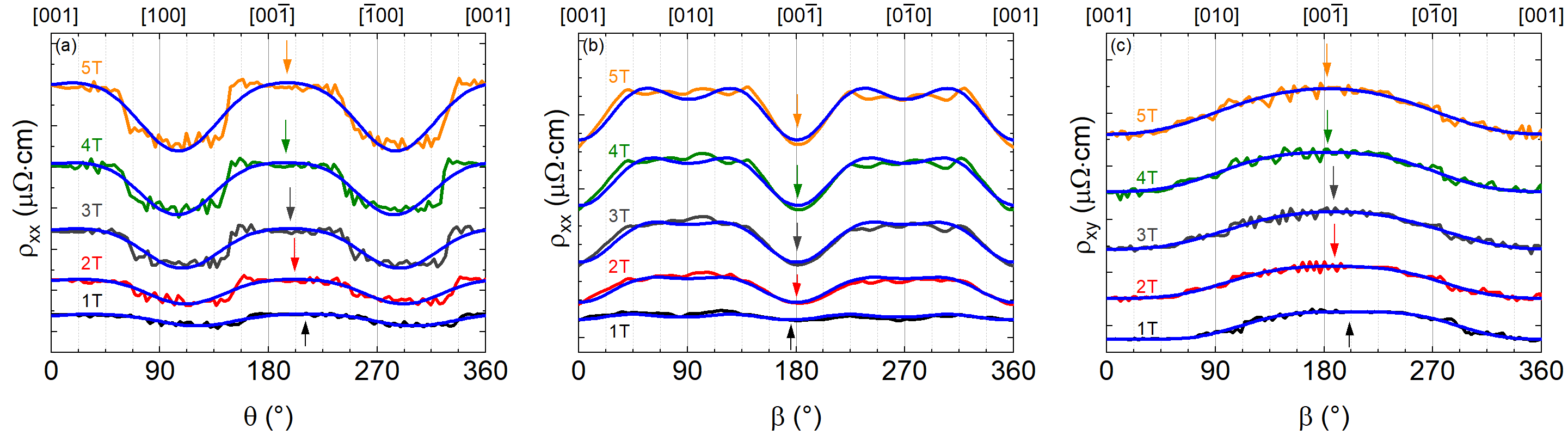}%
	\caption{\label{fig-admr}Angle-dependent magnetoresistance $\rho_{xx}$ (with differing constant offsets for various applied constant magnetic fields) for tilt from the sample normal to the in-plane direction parallel to the current (a) and from normal to in-plane perpendicular to current (b), as well as Hall resistivity $\rho_{xy}$ for the latter tilt direction (c). Modelling (see text) is shown by blue lines. Arrows indicate the orientation of the symmetry axes.}
\end{figure*}

For measuring the angle-dependent MR in a constant magnetic field $1$\,T$\leq\mu_0H\leq5$\,T the magnetization was first aligned parallel to the external field at 5\,T in an out-of-plane direction. Afterwards the sample is rotated in the plane parallel ($\theta$) and perpendicular ($\beta$) to the current measuring $\rho_{xx}$ and $\rho_{xy}$.

\begin{figure}[b!]
	\includegraphics[width=0.45\textwidth]{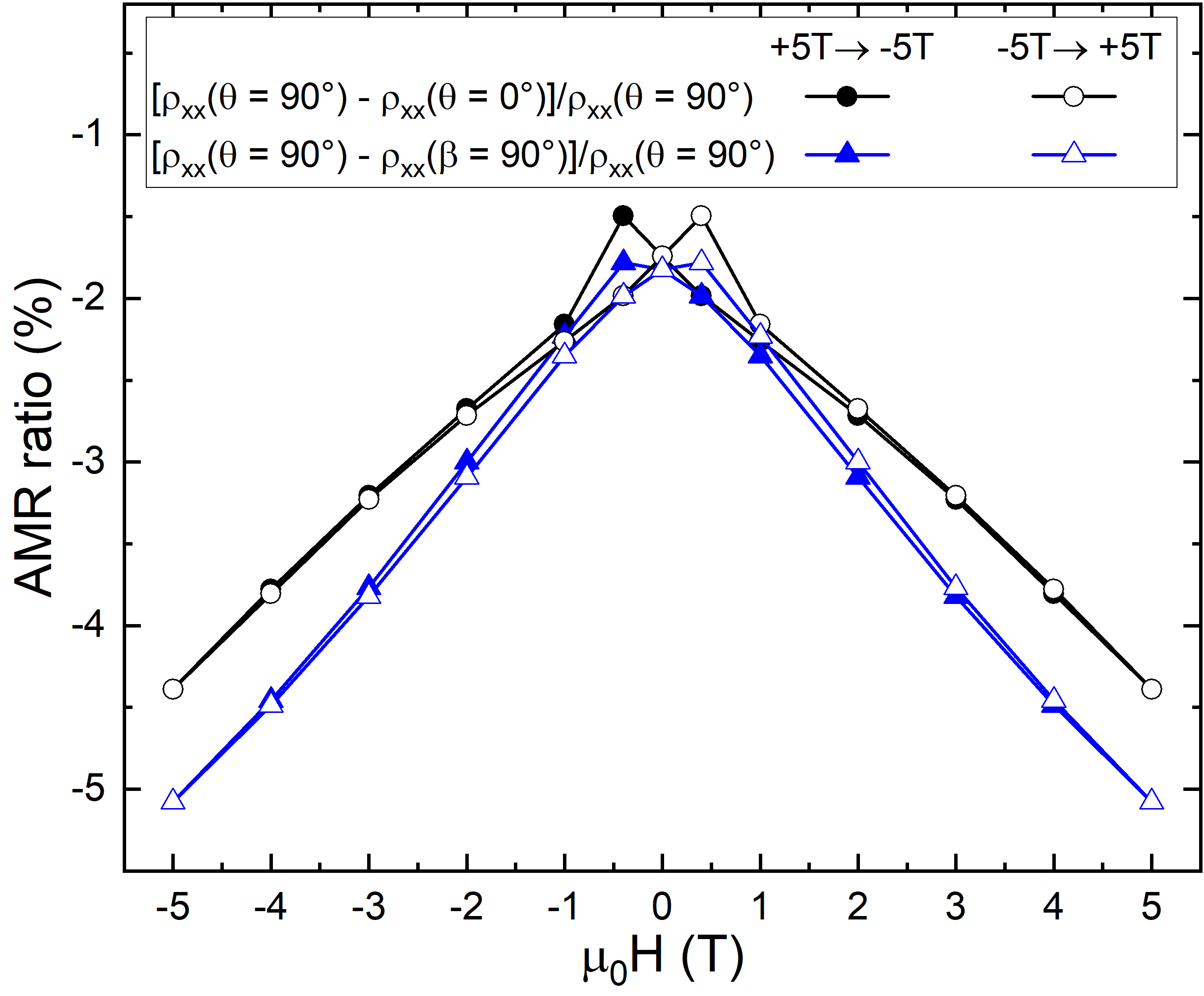}%
	\caption{\label{fig-Anisotric-Magnetoresistance}AMR ratio  (black) and (blue) of a [(SRO)$_5$/(SIO)$_2$)]$_{10}$ film at 10\,K.}
\end{figure}

For a magnetic field rotation perpendicular to the current direction (see Fig. \ref{fig-admr}(b) and (c)) a shift of the symmetry axis of the magnetization in the MR and Hall resistivity between $\mu_0H=1$\,T and $\mu_0H=5$\,T can be observed. The theoretical description of the angle-dependent magnetotransport properties in a constant magnetic field was done by a model introduced by Limmer \textit{et al.} \cite{Limmer2008}. At high fields $\mu_0H\geq4$\,T the symmetry axis aligned parallel to the out-of-plane direction of the sample, whereas it is tilted up to $20^\circ$ from out-of-plane for low fields. Comparing both tilt directions we observe only a small variation of the angle-dependent MR for the magnetic field rotation perpendicular to the current in contrast to the larger variation for tilting parallel to the current. 
The positive transversal magnetoresistance (TMR) which is measured for field rotation perpendicular to the current (cf. Fig. \ref{fig-theta-beta-angle-dependence-MR-with-Fit}(e)-(h)) displays only a small dependence on the magnetic field for in-plane and out-of-plane orientation. For $\theta=90^\circ$ magnetic and electric fields are in line, i.e. for this configuration the data represent the LMR, which is negative in our (SROSIO)$_{10}$ heterostructure for magnetic fields $|\mu_0 H|\geq 2$\,T.

From the MR curves at $\theta = 0^\circ$, $\theta = 90^\circ$ and $\beta = 90^\circ$ one can calculate the anisotropic magnetoresistance (AMR) using the definition from \cite{Limmer2006}. The field-dependence of the AMR at 10\,K is shown in Fig. \ref{fig-Anisotric-Magnetoresistance}. For both, out-of-plane and in-plane, the AMR is negative, whereas the in-plane AMR is larger than the out-of-plane. 
The in-plane AMR specifies the ratio between the LMR ($\theta=90^\circ$) and the in-plane TMR ($\beta=90^\circ$) while the out-of-plane AMR is given by the ratio of the LMR and out-of-plane TMR ($\theta=0^\circ$). Both the in-plane and out-of-plane TMR are larger than the LMR as indicated by the negative AMR ratio. The difference between both AMR ratios is related to the difference between the in-plane and out-of-plane TMR, This may arise from a tetragonal distortion of the unit cell.

\section{Conclusion}

Heterostructures of the bilayer SROSIO with various repetitions were grown on (001) oriented STO substrates. The strain transfer from the substrate through all layers of the heterostructure was checked by reciprocal space maps around ($l03$) and ($l04$) substrate peaks (see Fig. \ref{fig-XRD-XRR-RSM}(c) and Fig. S4 of Supplemental Material \cite{Supplemental}). Comparison of out-of-plane and in-plane magnetization indicates out-of-plane anisotropy. The FM ordering temperature $T_C$ and saturation magnetization for all heterostructures is reduced compared to pure SRO reference, stronger for less repetitions of bilayer. 

DFT+$U$+SOC calculations for a (SRO)$_5$/(SIO)$_2$) SL confirm the itinerant ferromagnetic state of the Ruthenate layers and moderate electronic reconstruction, leading to a negative AHE. They also show the crossing of several bands at the Fermi edge, justifying the application of the phenomenological two-band model for the description of magnetoresistance and Hall effect data.

Hall effect measurements at various temperatures with out-of-plane oriented magnetic field show distinct anomalies. Using a two channel anomalous Hall effect model we found in our symmetric [(SrRuO$_3$)$_5$/(SrIrO$_3$)$_2$)]$_{k}$ heterostructures that these anomalies do not originate from two independent spin-polarized conduction channels represented by anomalous Hall effects with different sign. Thus, we attribute these additional contributions to the topological Hall effect resulting from skyrmions in contrast to the asymmetric SrRuO$_3$ based heterostructures in \cite{Wysocki2020}, for which the total Hall effect could be modelled within the two-channel model. Temperature-dependent magnetoresistance with out-of-plane oriented magnetic field displays a crossover upon cooling below 60\,K from positive to negative MR. To fit the MR we introduce a two channel MR model, comparable to the two channel anomalous Hall effect model with similar coercitive fields.

Angle-dependent Hall effect measurements were performed for tilting the magnetic field from out-of-plane to in-plane, for both configurations perpendicular and parallel to the current $j$. The ordinary Hall constant $R_0$ and the shape of the Hall resistance show the same angle-dependence for both tilt directions. Adjusting the shape of the Hall effect by the two channel model we found an additional contribution $\Delta\rho_{\text{Hall}}$ for a large range of magnetic field orientation with respect to the film surface. Note, that such $\Delta\rho_{\text{Hall}}$ is not present in pure SRO thin films and therefore must be related to the interfaces in the heterostructures. We interpret it as THE due to non-coplanar magnetic structures like skyrmions of size $r_{sk} \approx 20$~nm. Analyzing the field dependence of $\Delta\rho_{\text{Hall}}$ yields, that these skyrmions are only stable for tilting the field from the sample normal to in-plane up to a critical angle of about $45^\circ$. Moreover we found for field-angles close to an in-plane orientation a peak-like anomaly in the Hall effect. This could be caused by a non-trivial spin structure different than N{\'{e}}el-type skyrmions due to the orientation of the magnetic moments parallel to the film plane. Possible candidates could be magnetic bimerons.

To further support the THE origin of $\Delta\rho_{\text{Hall}}$ in our samples we exclude alternative interpretations. For a defect-induced origin of the anomalies in the transverse resistivity (e.g. Ru vacancies \cite{Kan2018a}) or the existence of different magnetic domains \cite{Kan2018} $\Delta\rho_{\text{Hall}}$ reaches its maximum value close to the temperature of the AHE sign change. This is not the case in our thin films (c.f. Fig.\ref{fig-T-dependence-deltaRho}(a)) locating the maximum of the additional contribution at lower temperatures. Manipulation of the magnetic properties of the thin film resulting in a spatial variation of the magnetic and electronic structure could be another possible explanation. In a single SRO film this was realised through partial relaxation \cite{Miao2020}. Originating from such magnetic inhomogeneities both the shape of $\rho_{AHE}$ and the measured value of $\Delta\rho_{\text{Hall}}$ displays a negligible field-angle dependence. From reciprocal space maps of our sample (see Fig. \ref{fig-XRD-XRR-RSM}(c) and Fig. S4 of Supplemental Material \cite{Supplemental}) no evidence of structural relaxation could be observed. Furthermore, both $\rho_{AHE}$ and $\Delta\rho_{\text{Hall}}$ display a strong angular dependence. Therefore the emergence of anomalies in the Hall effect due to a (partial) relaxation induced inhomogenous magnetic structure is unlikely. In multilayers based on a symmetric trilayer of SIO (2 u.c.)-SRO (6 u.c.)-SIO (2 u.c.) the formation of skyrmions is unfavorable due to the symmetric boundary conditions in all SRO layers \cite{Yang2021}. The magnetization of the heterostructure with several repetitions of the trilayer indicates different Curie temperatures. This was interpreted as a variation of the magnetic properties between individual SRO layers, resulting in different magnetic and electronic structures for each SRO layer. For our [(SrRuO$_3$)$_5$/(SrIrO$_3$)$_2$)]$_{k}$ heterostructures there is no evidence of multiple $T_C$ values in the temperature-dependent magnetization (see Supplemental Material \cite{Supplemental}). Thus, there is no hint for an epitaxial strain induced variation of the magnetic properties of the individual SRO layers. The difference between our bilayer structure SRO (5 u.c.)-SIO (2 u.c.) and the SIO (2 u.c.)-SRO (6 u.c.)-SIO (2 u.c.) trilayer lies in the lower interface of the very first SRO layer, which is in our case between SRO and the STO substrate. Accordingly, the DMI from both interfaces of this layer does not cancel and the skyrmion formation becomes possible.

MR was measured for two tilt directions of the magnetic field, both parallel and perpendicular to the current. At fixed tilt angles the MR undergoes a crossover from positive to negative when inclining the magnetic field towards the direction of the current, whereas it stays positive for tilt direction perpendicular to the current.

The anisotropic magnetoresistance (AMR) is significantly larger (e.g. at 5\,T by about 18\,\%) for in-plane compared to out-of-plane orientation, consistent with isofield angular dependent measurements. The difference between both AMR ratios is related to the difference of in-plane and out-of-plane TMR, which could be caused by the tetragonal distortion of the unit cell. At low fields we observe a rotation of the magnetization direction from the out-of-plane direction which is reversed in larger fields. Altogether, our study on SROSIO heterostructures indicates a modification of the properties of pure SRO by the epitaxial SIO interfaces.  

\begin{acknowledgments}
	We thank S. Meir and V. Bruchmann-Bamberg for technical support and I. K{\'{e}}zsm{\'{a}}rki for useful discussions. This work was supported by the German Research Foundation (DFG) through the projects 107745057 (TRR80) and RO 5387/3-1. We acknowledge computational time at the Leibniz Rechenzentrum Garching, project pr87ro, and MagnitUDE (DFG Grant No. INST 20876/209-1 FUGG).
\end{acknowledgments}

%
%
\newpage
\setcounter{figure}{0}
\renewcommand{\thefigure}{S\arabic{figure}} 

\section*{Supplemental Material}
\author{Sven Esser}
\author{Jiongyao Wu}
\author{Sebastian Esser}
\author{Robert Gruhl}
\author{Anton Jesche}


\author{Vladimir Roddatis}

\author{Vasily Moshnyaga}
\author{Rossitza Pentcheva}


\author{Philipp Gegenwart}



\maketitle

Below, we provide additional information on the sample characterization and magnetotransport experiments. 

\subsection*{Sample characterization}

\begin{figure}[b!]
	\includegraphics[width=0.48\textwidth]{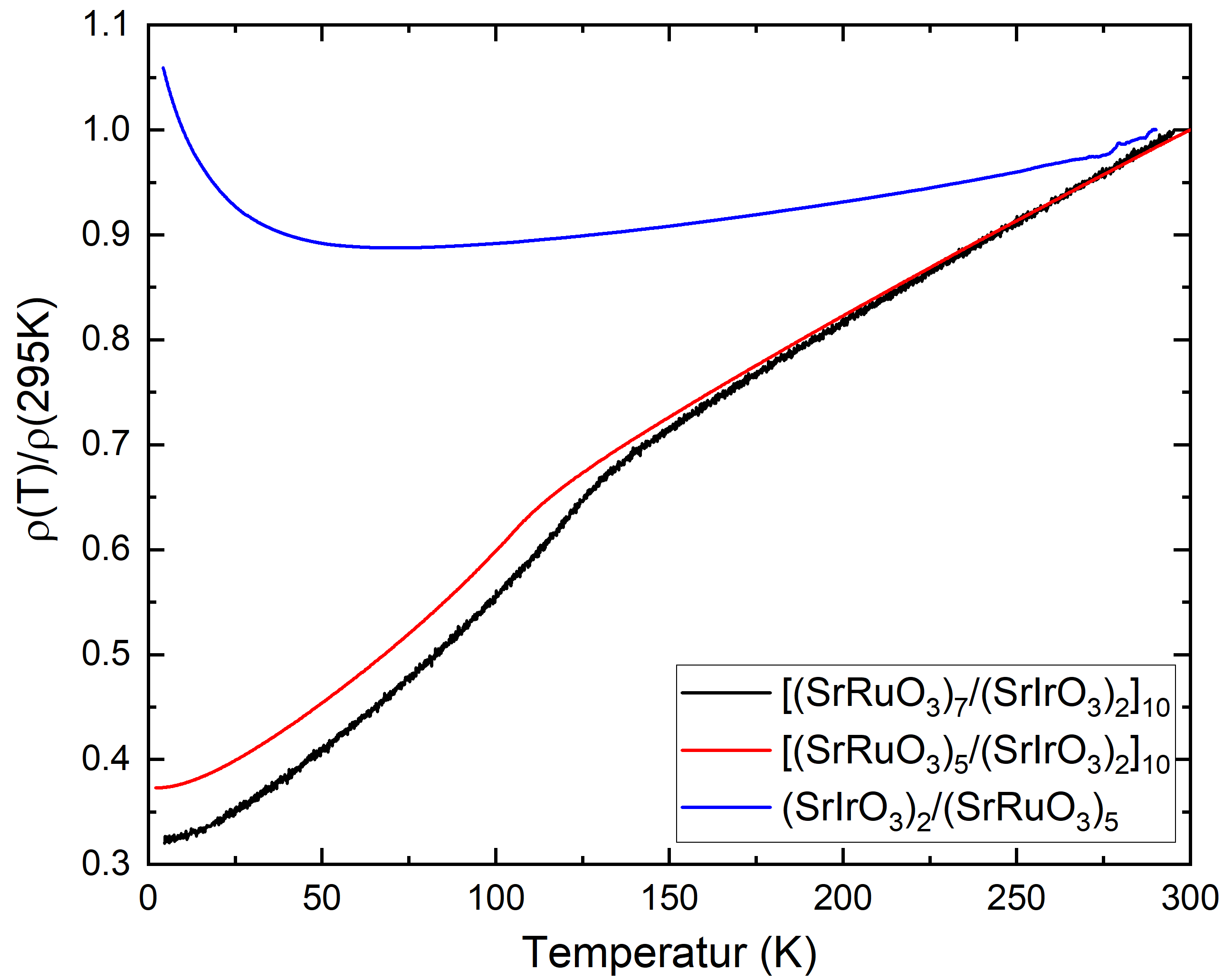}%
	\caption{\label{fig-R-vs-T-different-heterostrucutures}Normalized electrical resistivity versus temperature for different [(SRO)$_m$/(SIO)$_2$]$_k$ heterostructures indicated by the labels.}
\end{figure}

The temperature dependence of the electrical resistivity for [(SrRuO$_3$)$_5$/(SrIrO$_3$)$_2$)]$_{10}$ and  [(SrRuO$_3$)$_7$/(SrIrO$_3$)$_2$)]$_{10}$ heterostructures, as well as a [(SrIrO$_3$)$_2$)/(SrRuO$_3$)$_5$]$_{1}$ bilayer is displayed in Fig. \ref{fig-R-vs-T-different-heterostrucutures}. The resistivity of the heterostructures shows a clear kink at the Curie temperature $T_C$. In the low temperature regime finite-size effects such as weak localization \cite{Torre2001Sup} are observable for the bilayer.

\begin{figure}[t!]
	\includegraphics[width=0.5\textwidth]{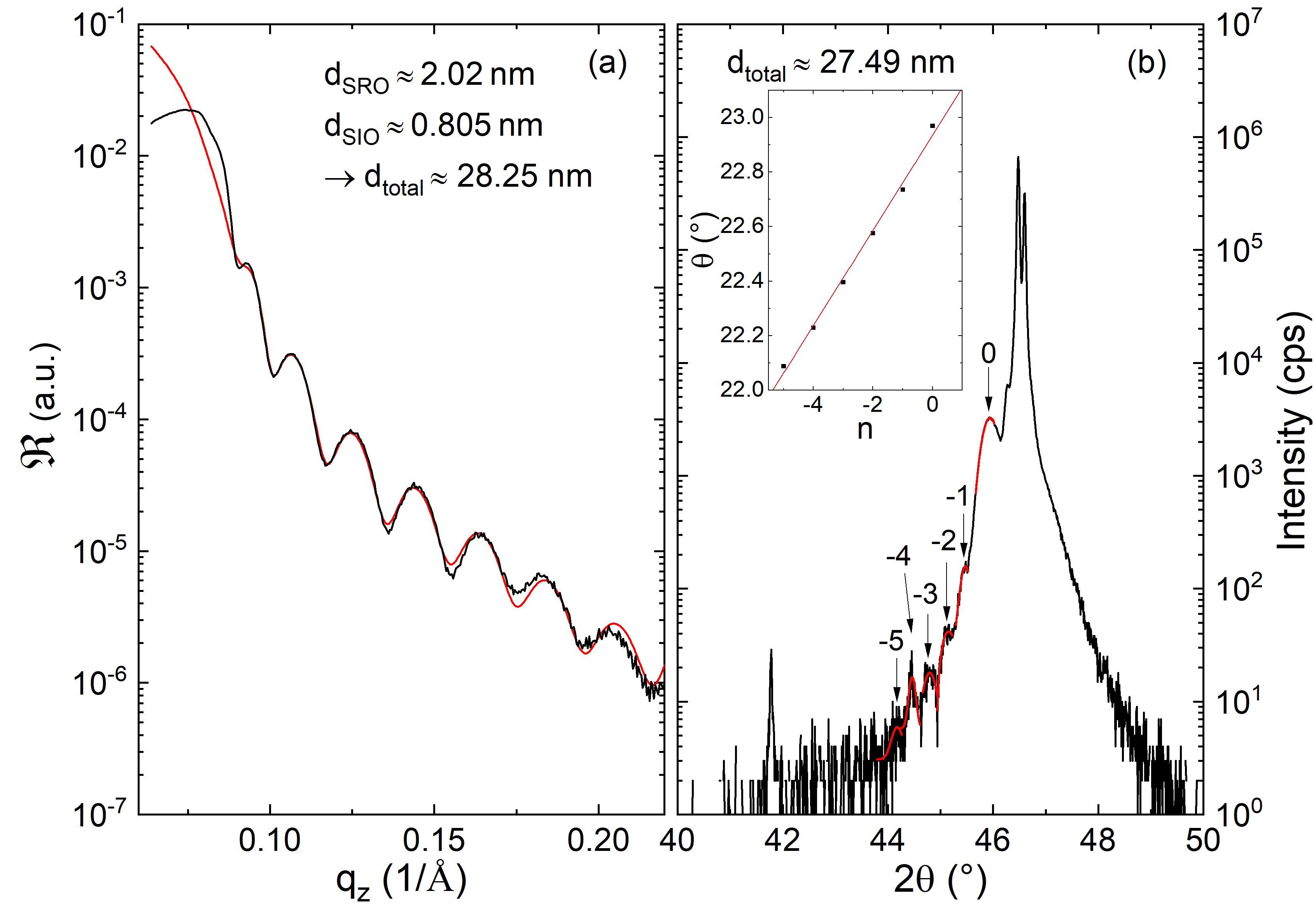}%
	\caption{\label{fig-laue-fringes-film-thickness-XRR-film-thickness-X'Pert}X-ray reflection (left) and diffraction (right) of [(SrRuO$_3$)$_5$/(SrIrO$_3$)$_2$)]$_{10}$. Red lines indicate simulations. Laue fringes are analyized in the inset.}
\end{figure}

\begin{figure}[b!]
	\includegraphics[width=0.5\textwidth]{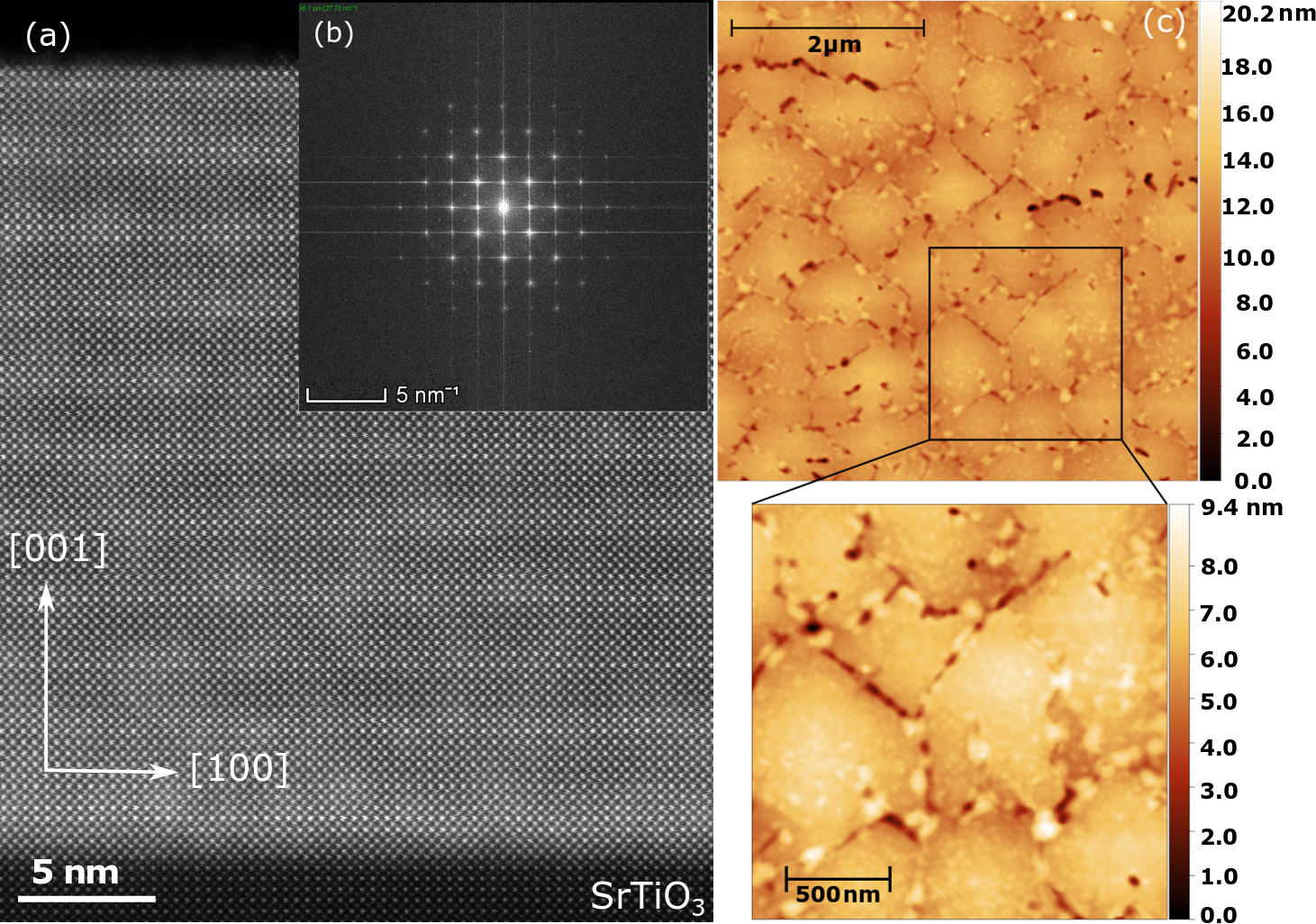}%
	\caption{\label{fig-TEM-AFM}(Color online) (a) HAADF HR-STEM image of a [(SRO)$_5$/(SIO)$_2$)]$_{10}$ heterostructure. (b) Fourier transformation of the (SROSIO)$_{10}$ area proves the (pseudo) cubic structure of the heterostructure. (c) AFM image of the film surface. Root-mean-square roughness of 1.1nm and $0.78$~nm on a 5x5~$\mu$m$^2$ and 2x2~$\mu$m$^2$ area, respectively proves the good quality of the grown films.}
\end{figure}

Fig. \ref{fig-laue-fringes-film-thickness-XRR-film-thickness-X'Pert} shows the X-ray reflection (XRR) and diffraction (XRD) pattern of a [(SrRuO$_3$)$_5$/(SrIrO$_3$)$_2$)]$_{10}$ heterostructure. In the XRD pattern Laue fringes are observable indicating the high sample quality. From the Laue fringes a total heterostructure film thickness of $27.49$~nm can be calculated. This agrees very well with the value estimate from the fit of the XRR data ($28.25$~nm). 

\begin{figure*}[t!]
	\includegraphics[width=1.0\textwidth]{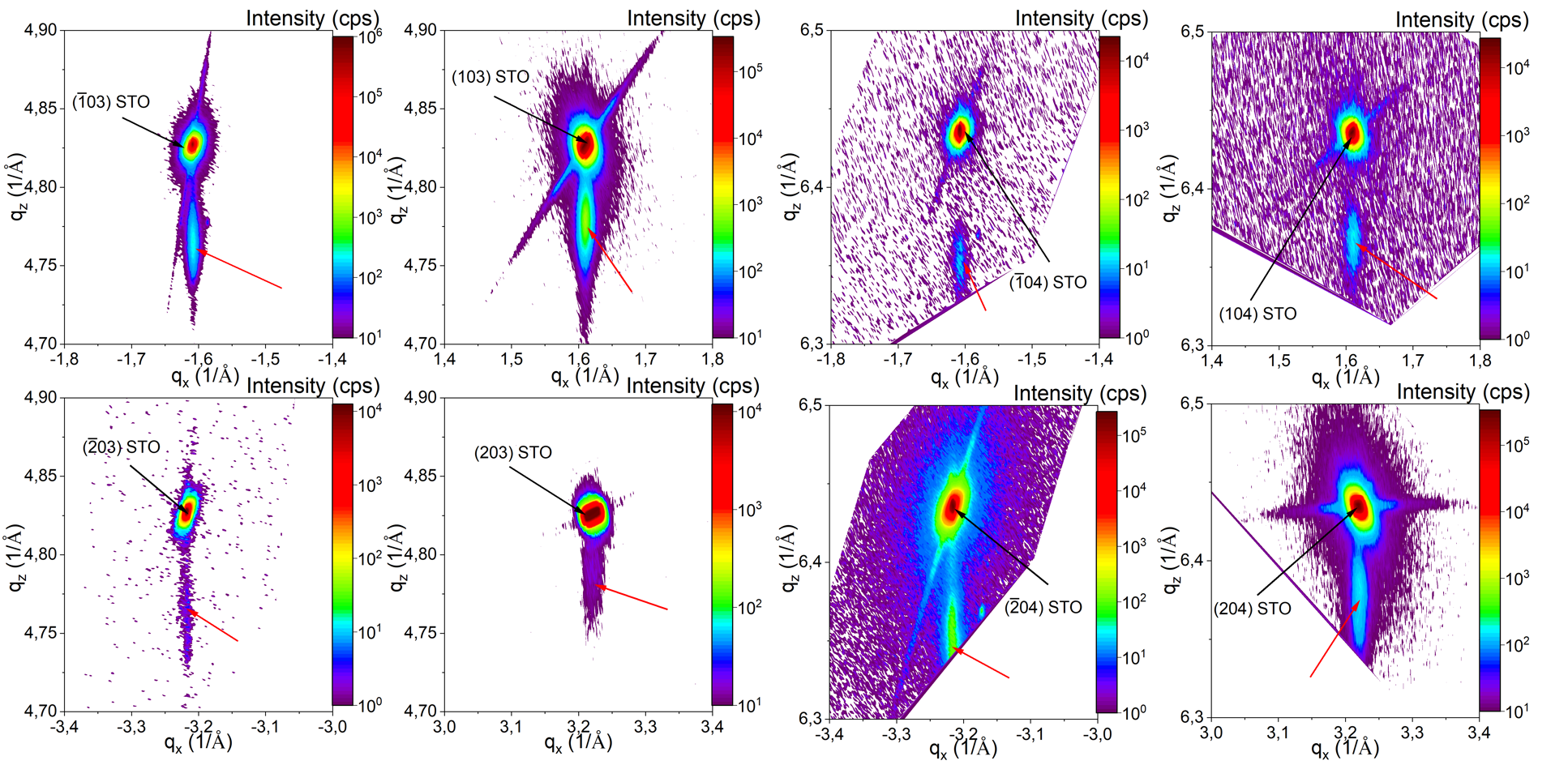}%
	\caption{\label{fig-RSM-L03-und-L04-peaks}Reciprocal space maps for a [(SRO)$_5$/(SIO)$_2$]$_{10}$ thin film around various ($h03$) and ($h04$) STO peaks (black arrows). The film peaks are marked by red arrows.}
\end{figure*} 

To verify the fully strained growth of the [(SrRuO$_3$)$_5$/(SrIrO$_3$)$_2$)]$_{10}$ thin film, reciprocal space maps (RSM) around different ($h03$) and ($h04$) peaks of the SrTiO$_3$ substrate where collected, shown in Fig. \ref{fig-RSM-L03-und-L04-peaks}. The position of the film peak is always located below the corresponding substrate peak. This indicates the fully strained state of the grown film \cite{Matsuno2015Sup}. From the peak position we can calculate the lattice constant for the total film in in-plane direction $a_{ip}=3.905\,\mathring{\text{A}}$ and out-of-plane $c_{oop}=3.949\,\mathring{\text{A}}$ equal to the distortion of the pseudo cubic unit cell with in-plane compression and out-of-plane elongation. The value of $c_{oop}$ agrees with the out-of-plane lattice constant from XRD which yields $c_{oop}=3.950\,\mathring{\text{A}}$.

\begin{figure*}[t!]
	\includegraphics[width=1.0\textwidth]{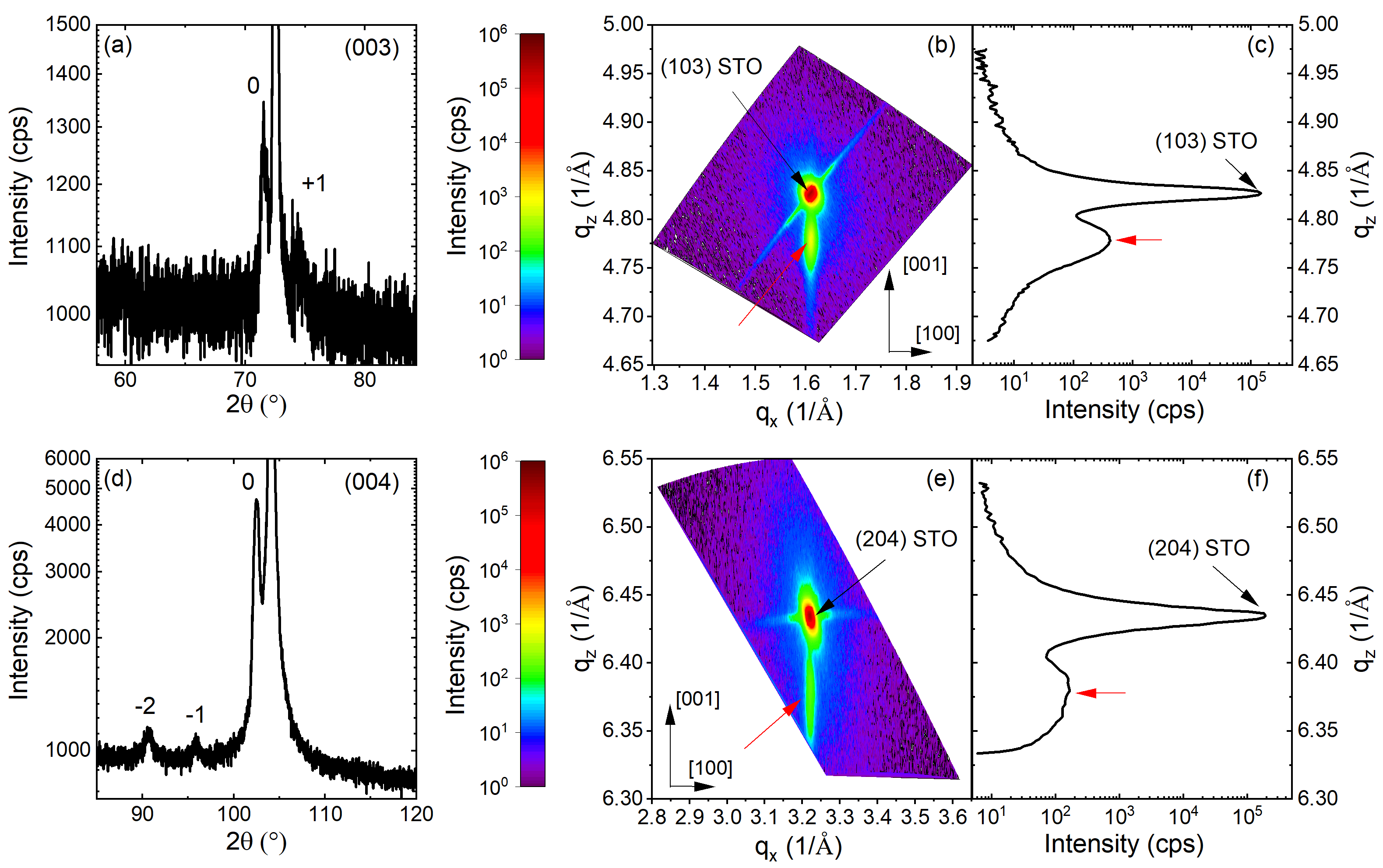}%
	\caption{\label{fig-RSM-closer}XRD and reciprocal space maps for a [(SRO)$_5$/(SIO)$_2$]$_{10}$ thin film. Small satellite peaks are visible beside the (003) film peak (a) and (004) film peak (d). In the reciprocal space map of the (103) (b) and (204) STO peak (e) the satellites are not
		visible due to their small intensities. The out-of-plane lattice constant of the film was determined from line profiles at $q_x=1.609\,\mathring{\text{A}}$ (c) and $q_x=3.218\,\mathring{\text{A}}$ (f). The film peaks are marked by red arrows.}
\end{figure*}

To further analyze the RSM, we firstly look closer on the (003)  (Fig.~\ref{fig-RSM-closer}(a))  and (004) (Fig.~\ref{fig-RSM-closer}(d)) peak of the XRD pattern. Satellites could be observed beside the main peaks.  The corresponding $q_z$ values and the intensity ratio between the satellites and the main peak are listed in Tab. \ref{table-xrd-satellites}. As the satellite peaks have much less intensity than the main peaks in RSM, here no satellites can be resolved.  In Fig.~\ref{fig-RSM-closer}(c) we show the line profile of the (103) RSM (Fig.~\ref{fig-RSM-closer}(b)) at $q_x = \frac{2\pi}{a_\text{STO}}\approx 1.61\,\mathring{\text{A}}^{-1}$. By determining the spot width we can obtain a distribution for the out-of-plane lattice constant $c_{oop}$ (see Tab. \ref{table-RSM-lineprofile}). Same is shown in Fig.~\ref{fig-RSM-closer}(f) for the (204) RSM (Fig.~\ref{fig-RSM-closer}(e)) at $q_x = 2\cdot\frac{2\pi}{a_\text{STO}}\approx 3.22\,\mathring{\text{A}}^{-1}$.  From this we can deduce that the out-of-plane lattice constant in our heterostructure has only a small variation less than $1\%$. 

\begin{table}[b!]
	\caption{\label{table-xrd-satellites} Analysis of satellite XRD peaks from Fig.~\ref{fig-RSM-closer}(a) and (d).}
	\begin{ruledtabular}
		\begin{tabular}{ccccc}
			peak   & satellite order $n$ & $2\theta$ ($^\circ$) & $q_z$ ($\mathring{\text{A}}^{-1}$) & intensity ratio $\frac{I(n)}{I(0)}$\\
			\hline
			(003)   &0	&71.59	&4.77	&1\\
			(003)   &+1 &74.41	&4.93	&0.48\\
			\hline
			(004)&-2	&90.70	&5.80	&0.04\\
			(004)&-1	&95.91	&6.08	&0.02\\
			(004)&0	&102.53	&6.36	&1\\
		\end{tabular}
	\end{ruledtabular}
\end{table}

The HR-STEM image in combination with its fast Fourier transformation (FFT) image (Fig \ref{fig-TEM-AFM}(a)-(b)) confirms the high structural quality of the heterostructure. No dislocations or other defects are visible. In the FFT image one can clearly observe the pseudo-cubic structure.

The AFM image of the SIOSRO thin film (Fig. \ref{fig-TEM-AFM}(c)) shows a smooth surface morphology. The calculated root-mean-square roughness $\text{RMS}=1.1$\,nm and $0.78$\,nm on a 5x5~$\mu$m$^2$ and 2x2~$\mu$m$^2$ area, respectively indicates the smooth growth of the $28.25$\,nm thick film, which is a further indication of sharp interfaces between the individual SIO/SRO layers.

\begin{table}[b!]
	\caption{\label{table-RSM-lineprofile}Analysis of line profiles from Fig.~\ref{fig-RSM-closer}(c) and (f), yielding the distribution of c-axis lattice constants with minimal, center and maximal values.}
	\begin{ruledtabular}
		\begin{tabular}{cccccc}
			& peak &  FWHM $\mathring{\text{A}}^{-1}$ & $c_\text{min}$ $\mathring{\text{A}}$ & $c_\text{max}$ $\mathring{\text{A}}$ & $c_\text{center}$ $\mathring{\text{A}}$ \\
			\hline
			SrTiO$_3$ & (103) & 0.007 & 3.902 & 3.908 & 3.905\\ 
			(SROSIO)$_{10}$ & (103) &  0.031 & 3.932 & 3.957 & 3.945\\ 
			\hline
			SrTiO$_3$ & (204) &  0.006 & 3.904 & 3.908 & 3.906\\ 
			(SROSIO)$_{10}$ & (204) &  0.058 & 3.923 & 3.959 & 3.941\\ 
		\end{tabular}
	\end{ruledtabular}
\end{table}

\begin{figure}[b!]
	\includegraphics[width=0.39\textwidth]{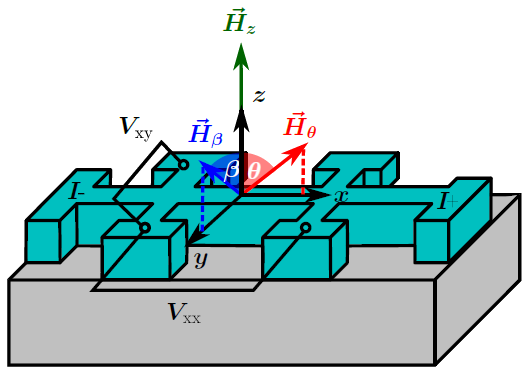}%
	\caption{\label{fig-defintition-tilt-angles}Orientation of the tilt angles $\beta$ and $\theta$ for the angle-dependent measurements.}
\end{figure}

\begin{figure}[b!]
	\includegraphics[width=0.5\textwidth]{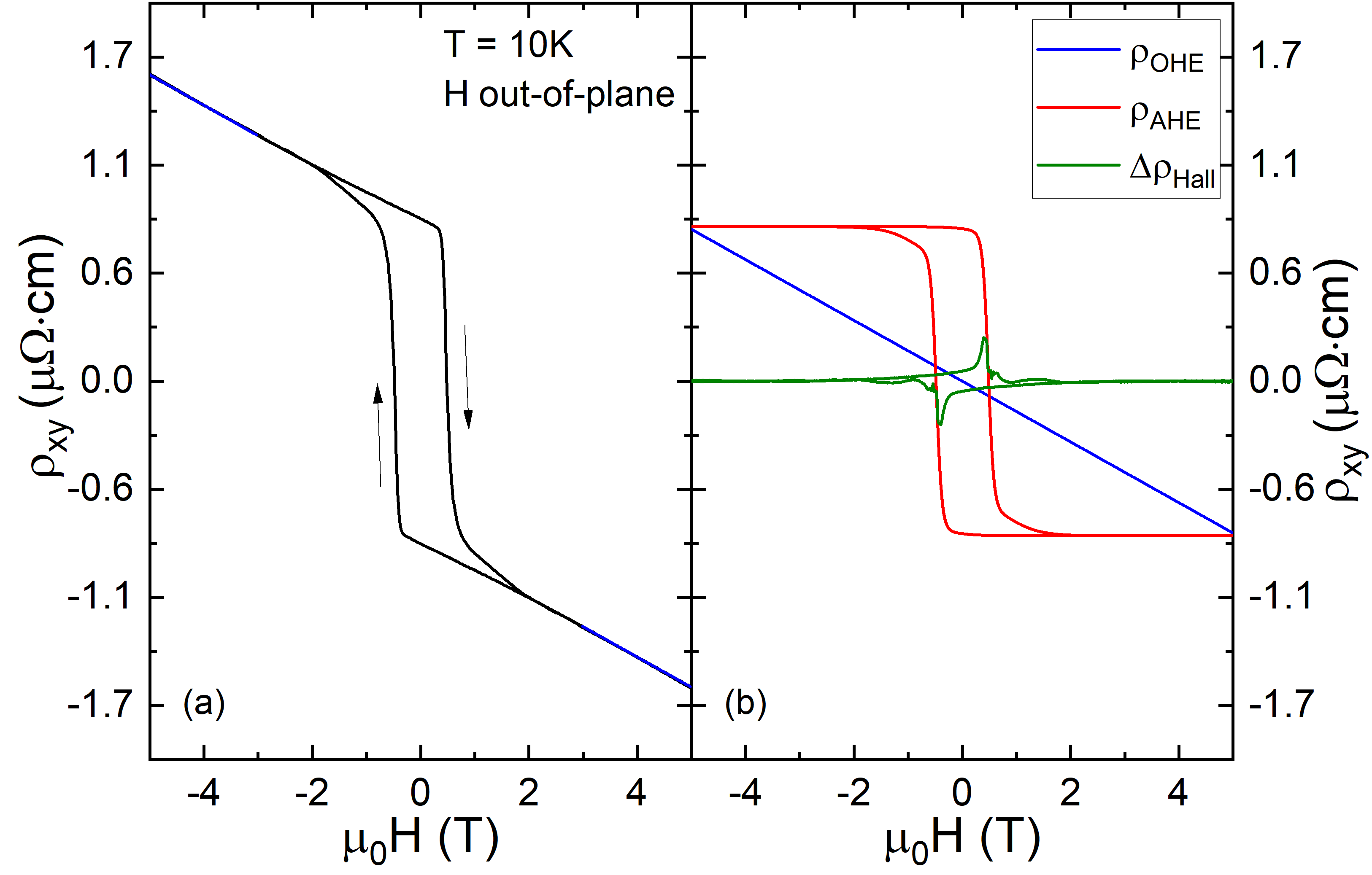}%
	\caption{\label{fig-determining-of-different-Hall-components}(a) Raw data for the Hall resistivity $\rho_{xy}$ (black curve) of a [(SRO)$_5$/(SIO)$_2$]$_{10}$ heterostructure at $T=10$\,K with out-of-plane oriented magnetic field. The ordinary Hall constant $R_0$ is identified by a linear fit in the high-field regime (blue curves). (b) Different contributions to the Hall resistivity $\rho_{xy}$: ordinary Hall effect $\rho_{OHE}$ (blue), anomalous Hall effect $\rho_{AHE}$ (red) and additional contribution $\Delta\rho_{\text{Hall}}$ (green).}
\end{figure}

\subsection*{Magnetotransport}

Angular dependent measurements were done for tilting the field from the sample normal to the in-plane direction in two different configurations, as indicated in Fig. \ref{fig-defintition-tilt-angles}. We define $\theta$ and $\beta$ as angles from the sample normal to the in-plane directions parallel and perpendicular to the current direction, respectively. Using this defintion $\theta$ describes the rotation of the magnetic field in the $xz$ plane and $\beta$ in the $yz$ plane. 

\begin{figure}[t!]
	\includegraphics[width=0.45\textwidth]{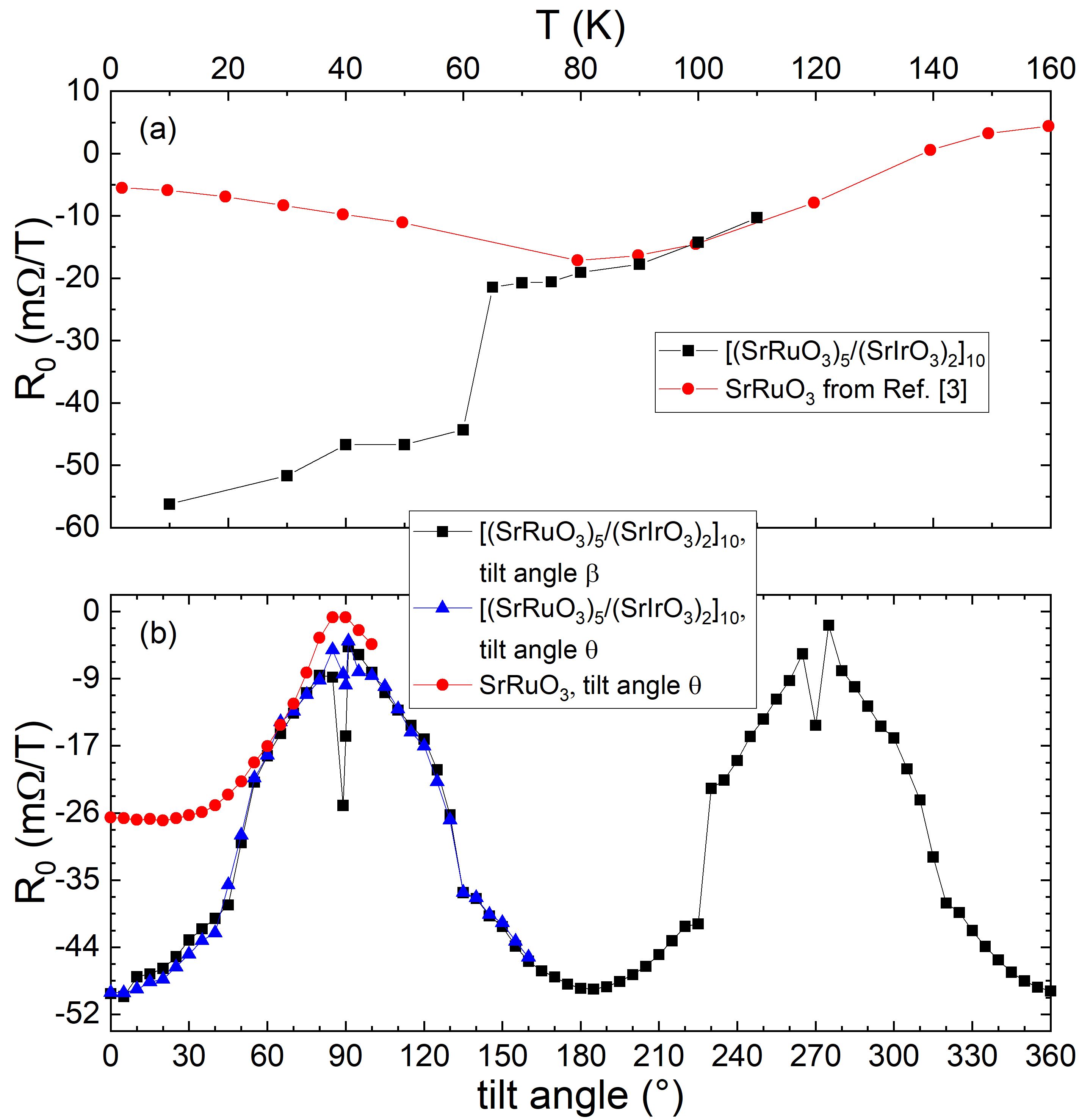}%
	\caption{\label{fig-ordinary-Hall-constant-temperature-angle-dependence_comparing-to-SRO}(a) Temperature-dependence of ordinary Hall constant $R_0$ of a [(SRO)$_5$/(SIO)$_2$]$_{10}$ heterostructure and a pure SrRuO$_3$ thin film from \cite{Srba2018Sup}. (b) Comparing angle-dependence of $R_0$ for magnetic field rotation for both tilting directions ($\beta, \theta$) for [(SRO)$_5$/(SIO)$_2$]$_{10}$ and for SrRuO$_3$ in $\theta$ direction at $T=10$\,K.}
\end{figure}

As shown in Fig. \ref{fig-determining-of-different-Hall-components}(a), the ordinary Hall constant $R_0$ is determined by a linear fit in the high-field regime and shows an electron-like character of the charge carriers for temperatures $T<T_C$. The temperature dependence of $R_0$ is shown in Fig. \ref{fig-ordinary-Hall-constant-temperature-angle-dependence_comparing-to-SRO}(a). Upon warming, a pronounced change occurs between 60\,K and 70\,K, leading to values which are closer to those for pure SrRuO$_3$ thin films \cite{Srba2018Sup}. 

The angular dependence of $R_0$ at 10\,K is presented in Fig. \ref{fig-ordinary-Hall-constant-temperature-angle-dependence_comparing-to-SRO}(b). Both tilt directions show a similar angle-dependence of $R_0$ with a cosine like shape. As there is no difference for the ordinary Hall constant between rotation in the $xz$ and $yz$ plane, we conclude that there is no dependence of the ordinary Hall effect on the tilt direction of the magnetic field. Compared to a SrRuO$_3$ film we observe for our heterostructure a more negative value of $R_0$ for tilt angles between $0^\circ$ (out-of-plane direction) and $60^\circ$, where there is a similar value for tilt angles between $60^\circ$ and $90^\circ$ (in-plane direction).

\begin{figure}[b!]
	\includegraphics[width=0.5\textwidth]{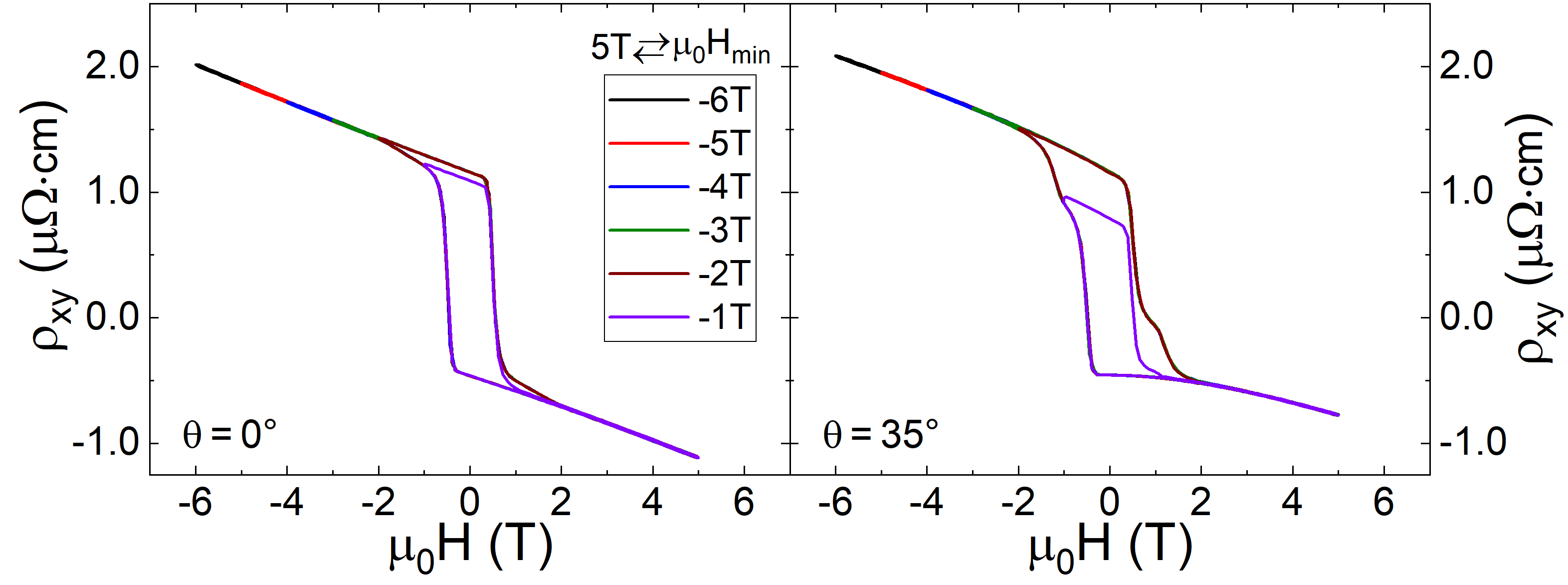}%
	\caption{\label{fig-minor-loops-asymmetric-theta-10K}Asymmetric field-loops for the angle-dependent Hall effect of a [(SRO)$_5$/(SIO)$_2$]$_{10}$ thin film.}
\end{figure}

\begin{figure}[t!]
	\includegraphics[width=0.5\textwidth]{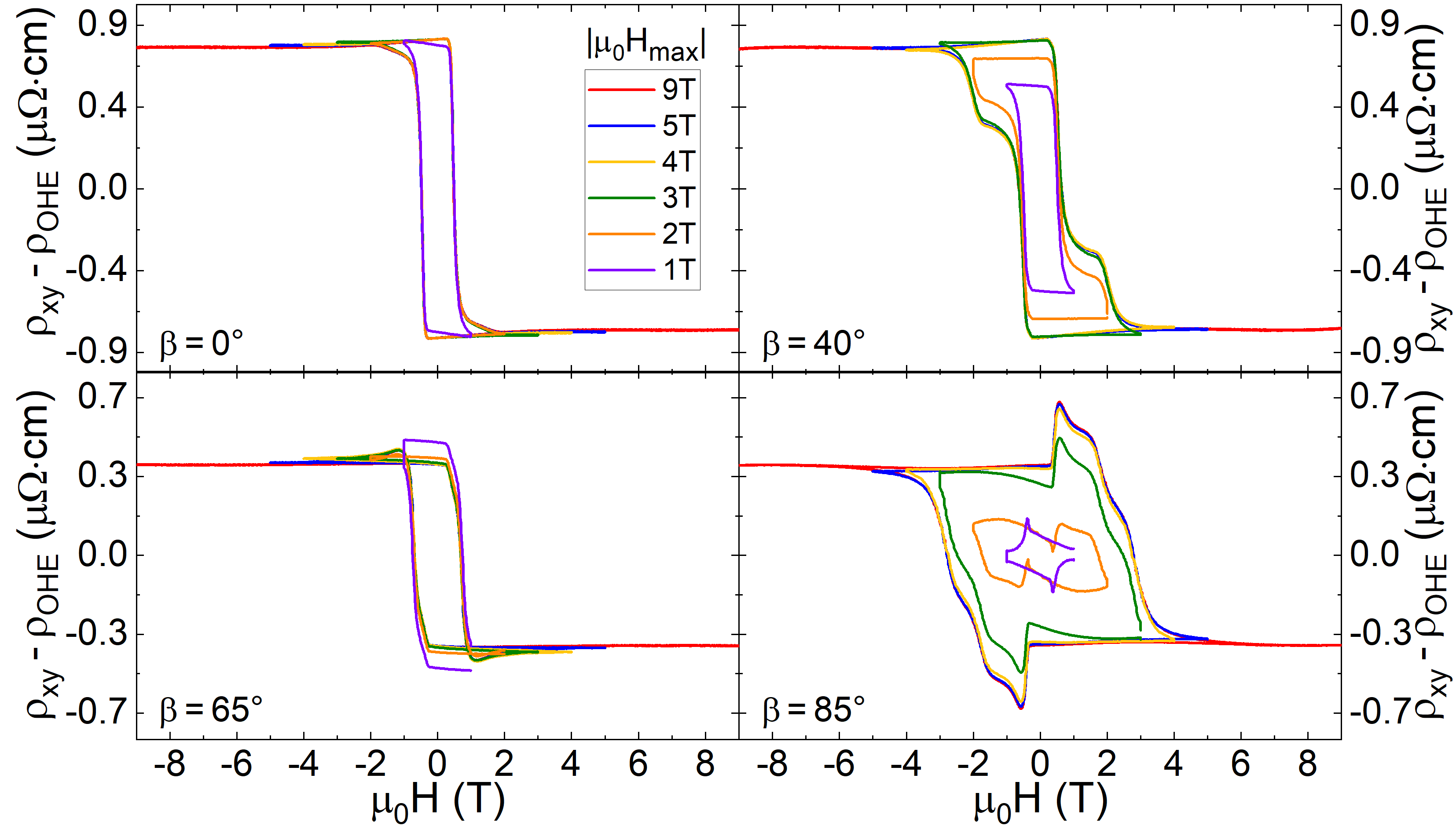}%
	\caption{\label{fig-minor-loops-symmetric-beta-10K}Symmetric field-loops for the angle-dependent Hall effect of a [(SRO)$_5$/(SIO)$_2$]$_{10}$ thin film. The ordinary Hall effect $\rho_{OHE}$ has been subtracted.}
\end{figure}

\begin{table}[b!]
	\caption{\label{table-T-dependence-Hall-Fit-Parameter} Temperature-dependence of coercitive fields $H_{c1,2}$ from two channel anomalous Hall effect fits for out-of-plane oriented magnetic field.}
	\begin{ruledtabular}
		\begin{tabular}{ccc}
			$T$ (K)   & $H_{c1}$ (T) & $H_{c2}$ (T)\\
			\hline
			10&	0.83&	0.48\\
			30&	0.84&	0.29\\
			50&	0.71&	0.22\\
			60&	0.60&	0.21\\
			65&	0.53&	0.18\\
			70&	0.45&	0.20\\
			80&	0.08&	0.16\\
			90&	0.08&	1.25\\
			100&	0.04&	0.01\\
			110&	0.03&	0.02\\
		\end{tabular}
	\end{ruledtabular}
\end{table}

\begin{figure}[b!]
	\includegraphics[width=0.5\textwidth]{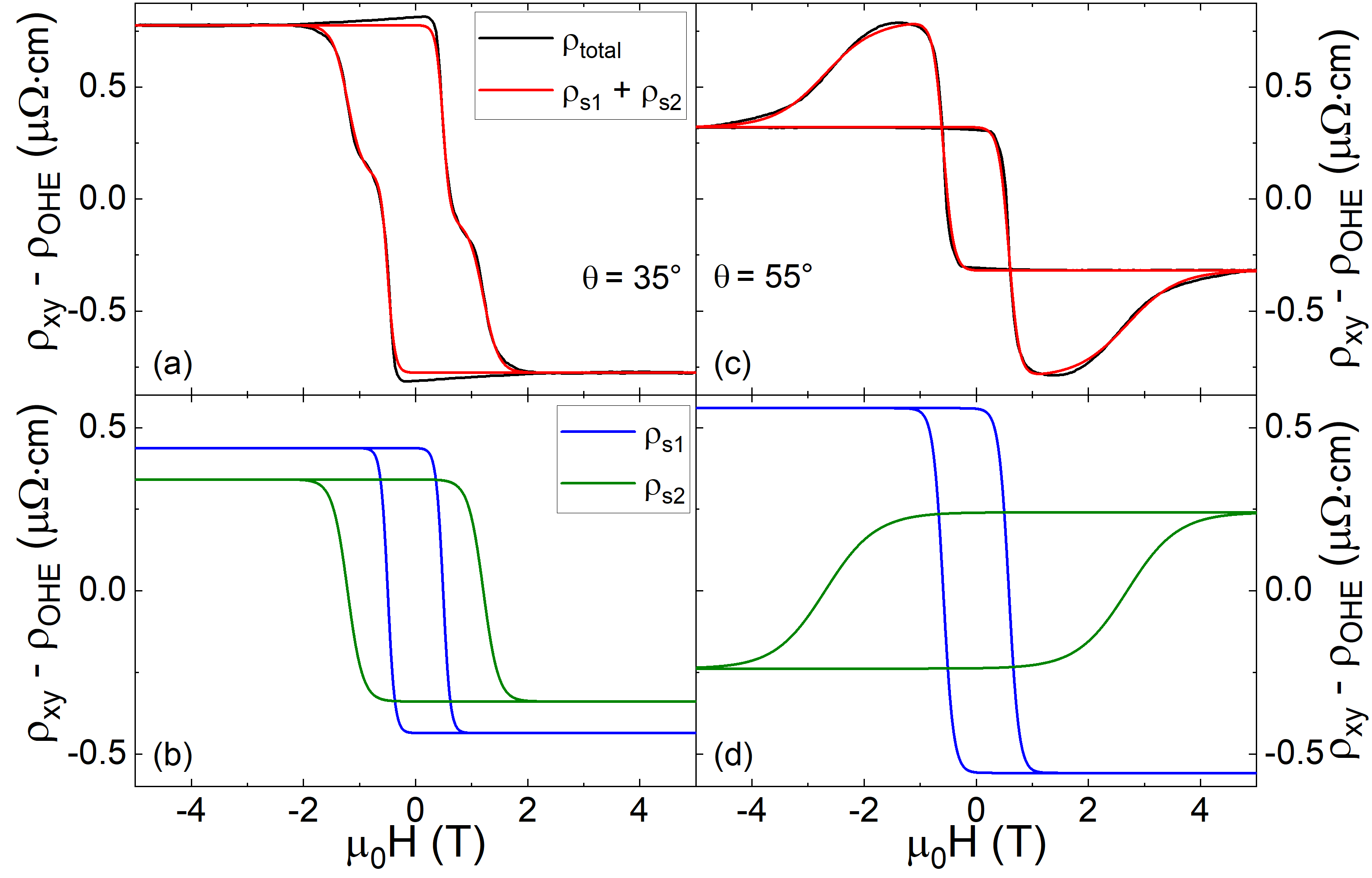}%
	\caption{\label{fig-10K-35deg-55deg-theta-Hall-seperated}Hall effect measurement (black) and  two channel fit (red) at $T=10$\,K for $\theta=35^\circ$ (a) and $\theta=55^\circ$ (c). The shape of the independent conduction channels $\rho_{s1}$ and $\rho_{s2}$ of the anomalous Hall effect $\rho_{AHE}$ for $\theta=35^\circ$ (b) and $\theta=55^\circ$ (d).}
\end{figure}

Symmetric field-loops for the Hall effect at 10\,K were performed for maximum magnetic fields 1\,T$\leq|\mu_0H_{\text{max}}|\leq9$\,T (see Fig. \ref{fig-minor-loops-symmetric-beta-10K}) and asymmetric field-loops for field sweeps from $5\,\text{T}$ to $\mu_0H_{\text{min}}$, with $-6\,\text{T}\leq\mu_0H_{\text{min}}\leq-1\,\text{T}$ (Fig.  \ref{fig-minor-loops-asymmetric-theta-10K}). As illustrated they show similar shapes for fields $|\mu_0H_{\text{max}}|\geq2$\,T. Considering the two channel anomalous Hall effect model introduced by Groenendijk \textit{et al.} \cite{Groenendijk2018Sup, Thiel2020Sup}, we point out that a maximum field of $|\mu_0H_{\text{max}}|=1$\,T is not able to switch rather than saturate both channels. Thus, we only see the contribution of one of the anomalous Hall channels. 

\begin{figure}[t!]
	\includegraphics[width=0.5\textwidth]{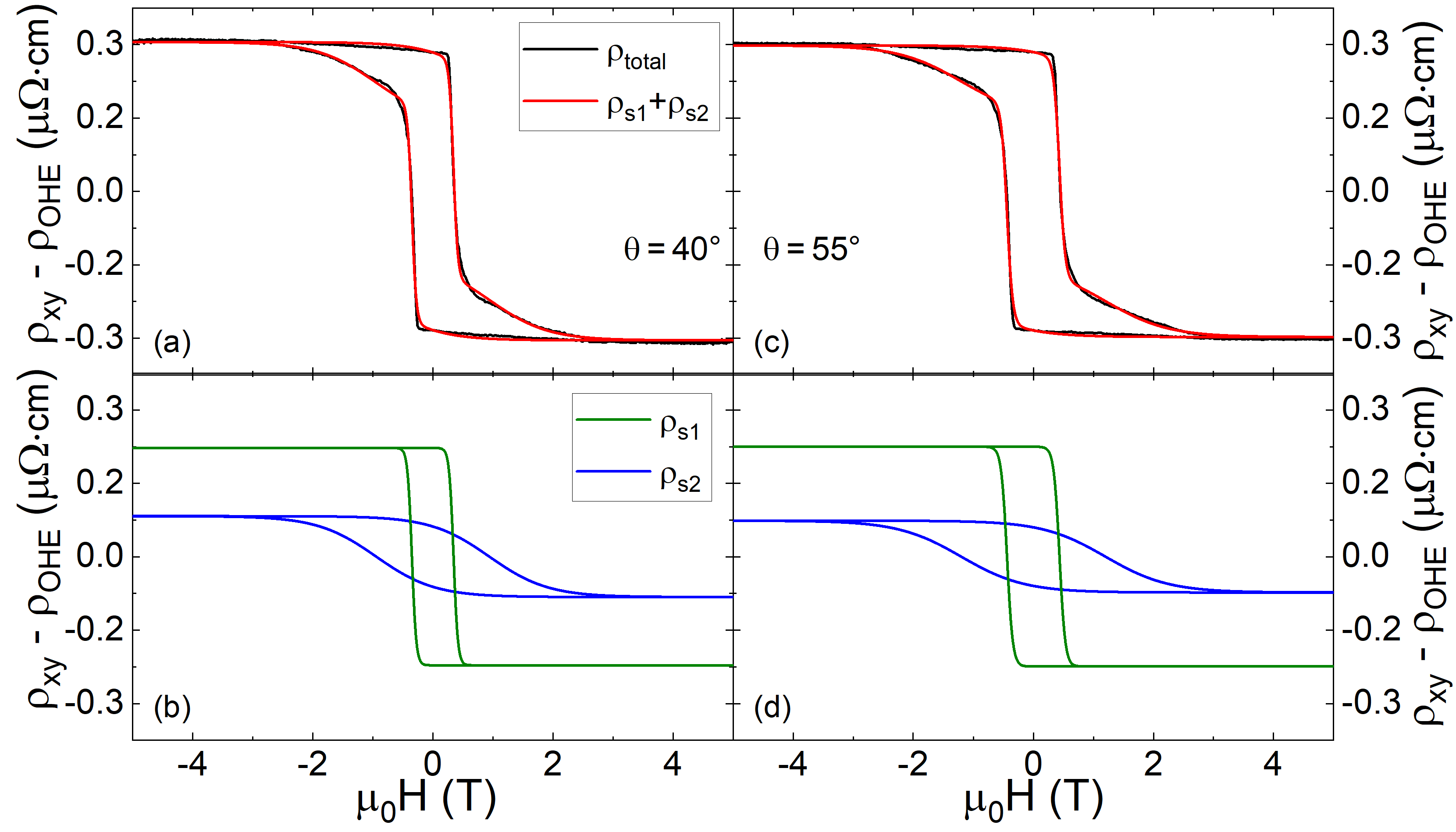}%
	\caption{\label{fig-10K-40deg-55deg-theta-Hall-seperated}Hall effect measurement (black) for a SrRuO$_3$ film and two channel fit (red) at $T=10$\,K for $\theta=40^\circ$ (a) and $\theta=55^\circ$ (c). The shape of the independent conduction channels $\rho_{s1}$ and $\rho_{s2}$ of the anomalous Hall effect $\rho_{AHE}$ for $\theta=40^\circ$ (b) and $\theta=55^\circ$ (d).}
\end{figure}

Comparing the angle-dependence of Hall effect (Fig. \ref{fig-10K-35deg-55deg-theta-Hall-seperated}) we find for the two channels of the anomalous Hall effect model the same sign at $\theta=35^\circ$ and opposite sign at $\theta=55^\circ$.

The angle-dependent Hall effect for a pure SRO thin film can also be modelled using the same two-channel anomalous Hall effect model as for our SROSIO heterostructures. As shown in Fig. \ref{fig-10K-40deg-55deg-theta-Hall-seperated} for $\theta=40^\circ$ and $\theta=55^\circ$ the Hall effect can be described without an additional 	contribution $\Delta\rho_{\text{Hall}}$. In contrast to our heterostructures we find for the SRO film that both conduction channels $\rho_{s1}$ and $\rho_{s2}$ of the anomalous Hall effect contribute with the same sign for all field-angles between 0 and $90^\circ$.

\begin{figure}[b!]
	\includegraphics[width=0.45\textwidth]{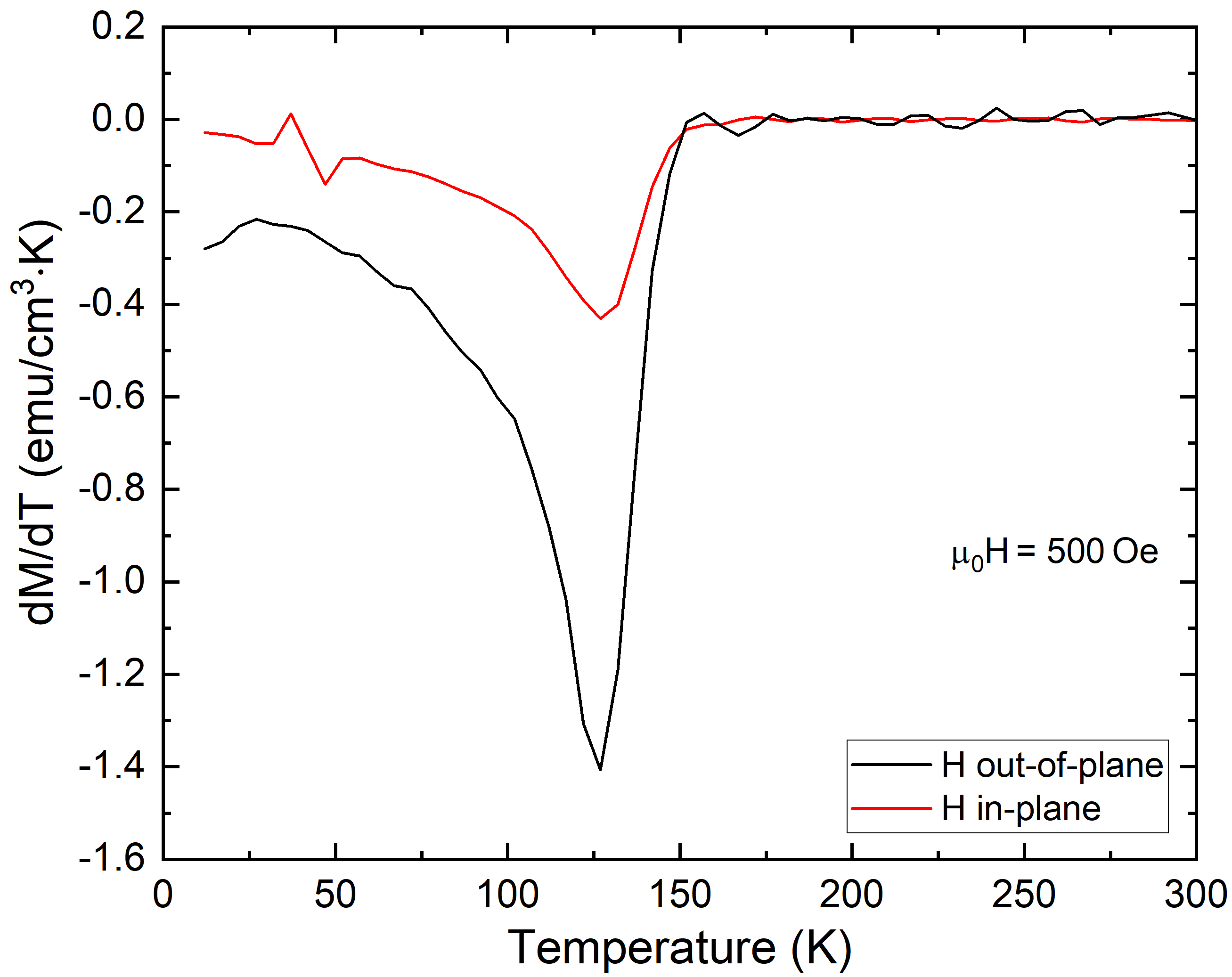}%
	\caption{\label{fig-M-vs-T-with-derivative}Derivative $dM/dT$ of the temperature dependent magnetization for a [(SRO)$_5$/(SIO)$_2$]$_{10}$ thin film in an out-of-plane (black) and in-plane (red) oriented magnetic field. The peak displays the occurance of a single $T_C$.}
\end{figure}

\begin{figure}[t!]
	\includegraphics[width=0.5\textwidth]{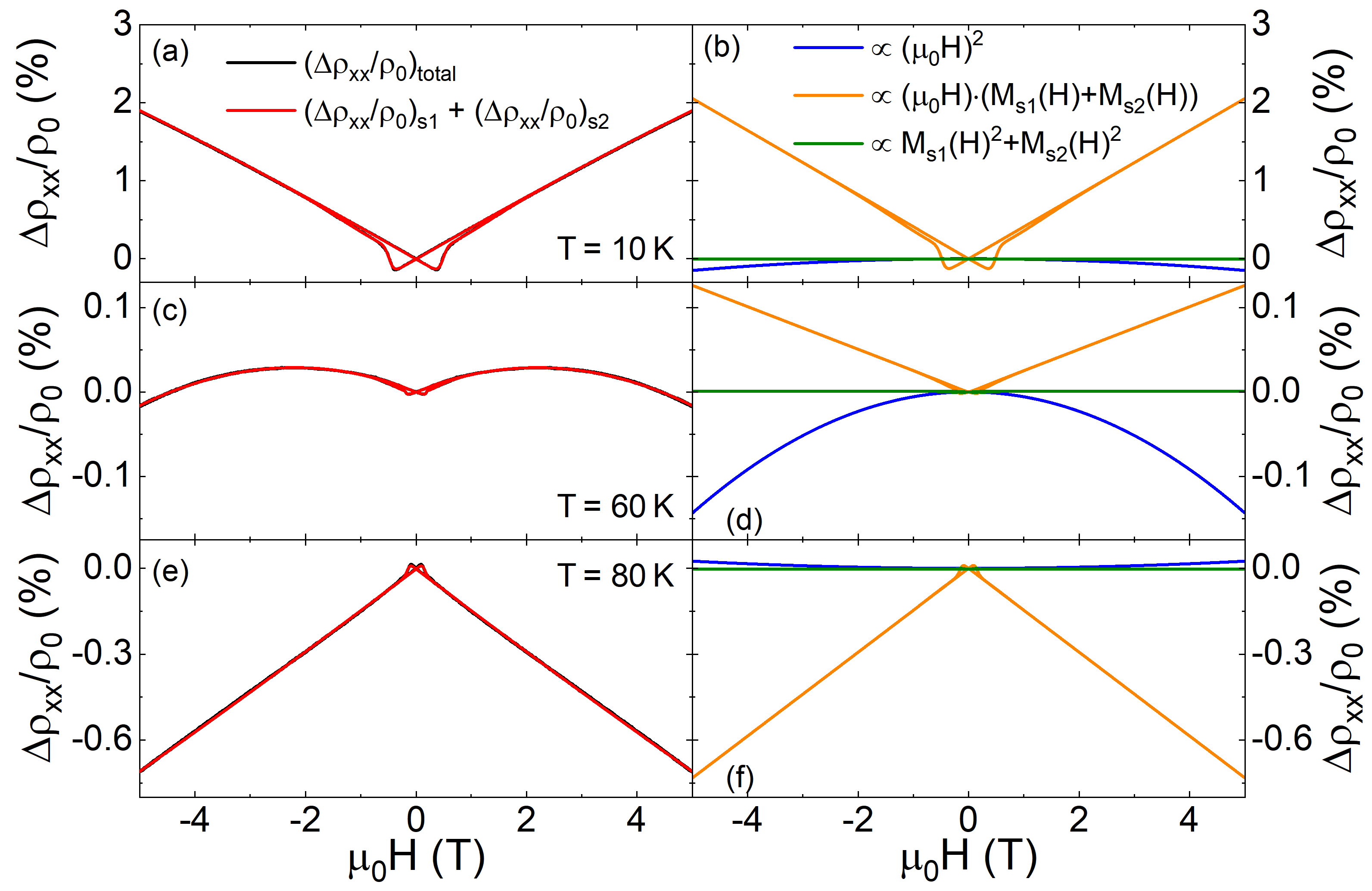}%
	\caption{\label{fig-10K-60K-80K-oop-MR-seperated}Temperature-dependent magnetoresistance with out-of-plane oriented magnetic field at $T=10$\,K (a), $T=60$\,K (c) and $T=80$\,K (e). Black lines show the measurement and red lines the two channel fit. Separated magnetoresistance fit by powers of magnetic field strength for $T=10$\,K (b), $T=60$\,K (d) and $T=80$\,K (f).}
\end{figure}

\begin{figure}[t!]
	\includegraphics[width=0.4\textwidth]{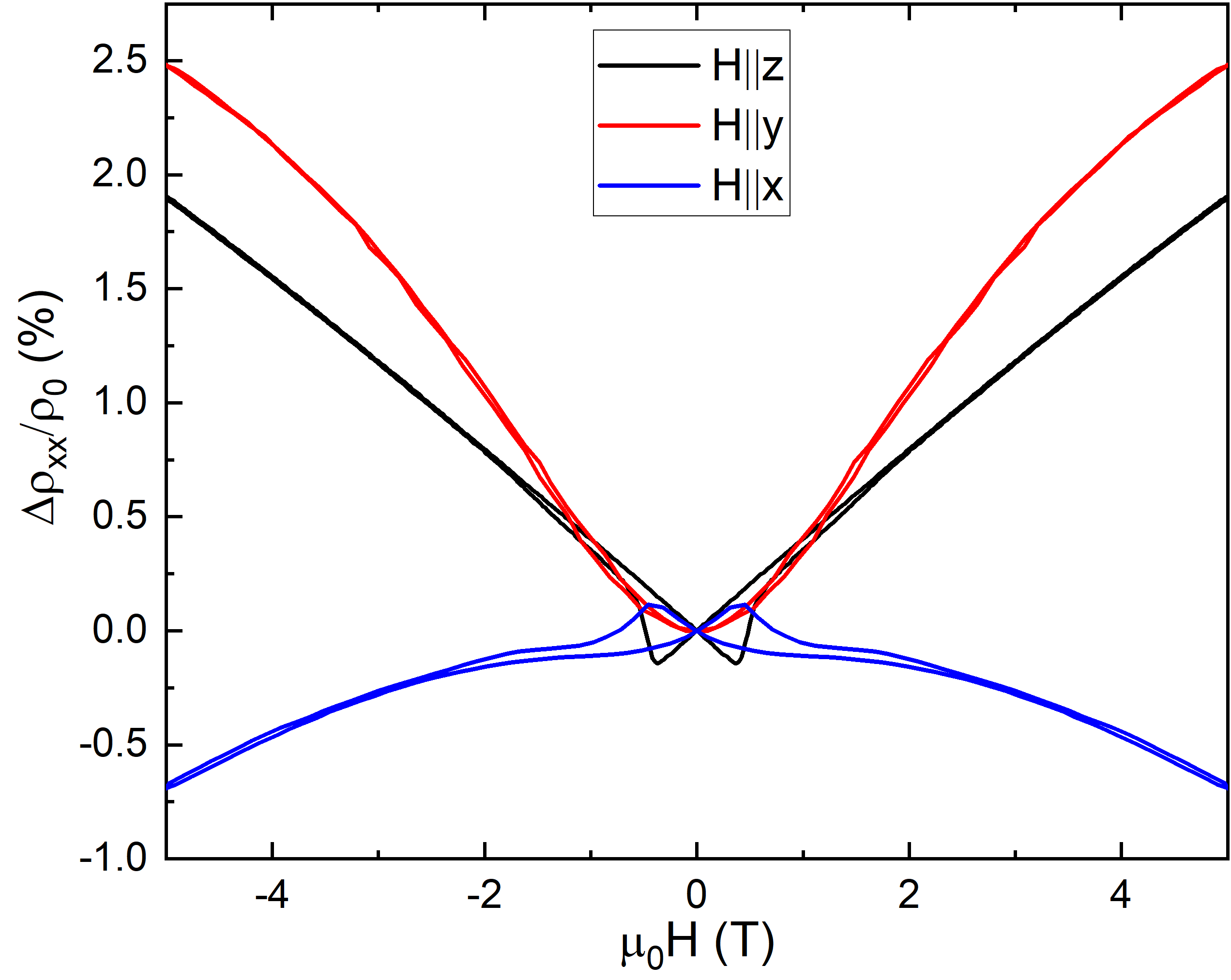}%
	\caption{\label{fig-10K-AMR}Magnetoresistance at $=10$\,K for magnetic field orientation out-of-plane and perpendicular to current direction ($H\parallel z$) and in-plane (perpendicular to current direction ($H\parallel y$) and parallel to current ($H\parallel x$)).}
\end{figure}

\begin{figure}[h!]
	\includegraphics[width=0.5\textwidth]{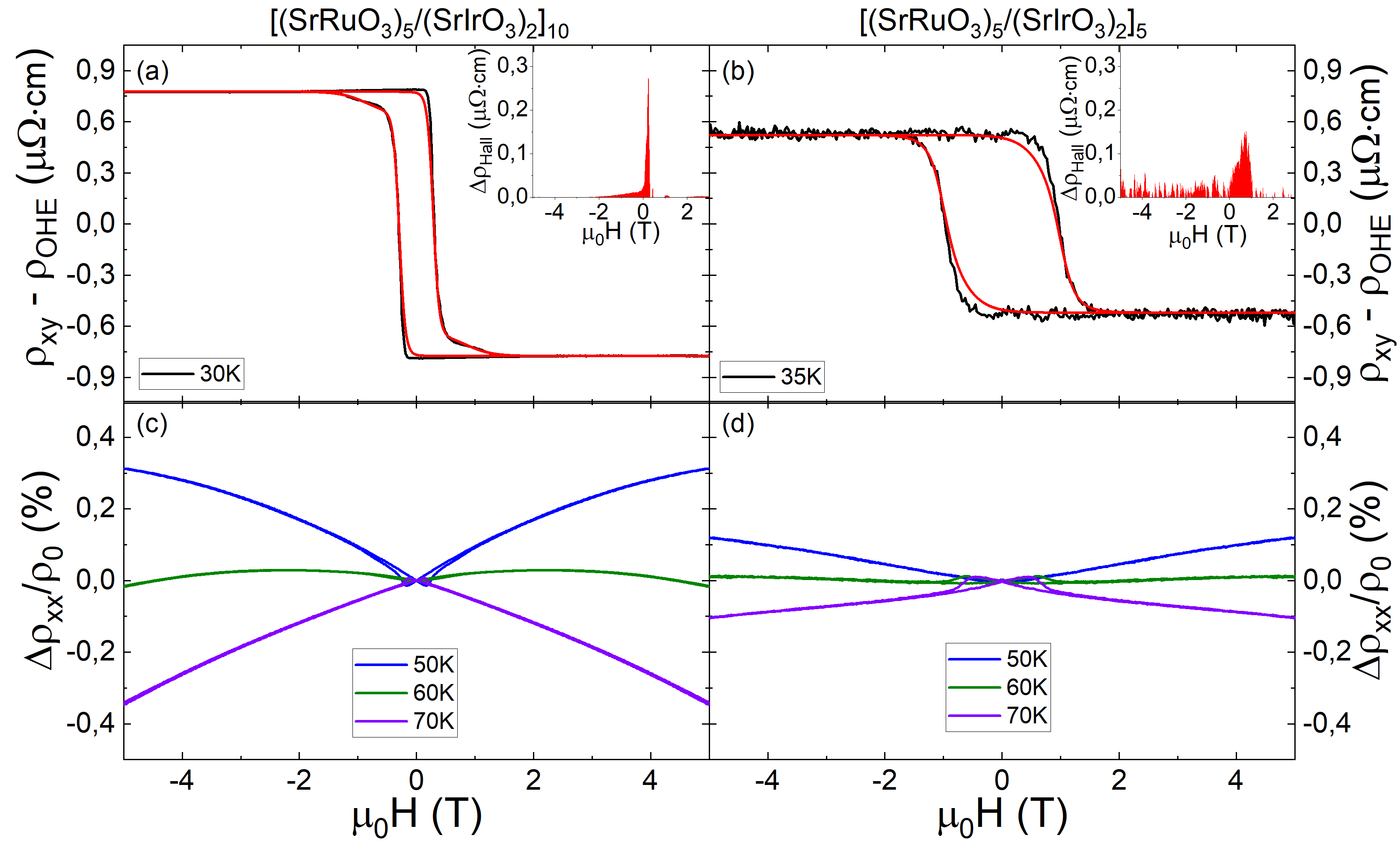}%
	\caption{\label{fig-T-dep-MR-Hall-K5-K10}Hall effect with out-of-plane oriented magnetic field of [(SRO)$_5$/(SIO)$_2$)]$_{k}$ heterostructures with $k=10$ (a) and $k=5$ (b) at 30 and 35~K, respectively. Black lines show measured data, while the red lines display the fits with the two channel model. The inset shows the additional contribution $\Delta\rho_{\text{Hall}}$. Note the similarity between the results on the two different heterostructures. The shape of the magnetoresistance for the $k=10$ (c) and $k=5$ (d) heterostructures is also similar and the sign change occurs in the same temperature interval.}
\end{figure}

The derivative of temperature-dependent magnetization curves for a [(SRO)$_5$/(SIO)$_2$)]$_{10}$ thin film in an in-plane and out-of-plane magnetic-field orientation (Fig. \ref{fig-M-vs-T-with-derivative}) is used to investigate the possible appearance of multiple ordering temperatures $T_C$ indicating a variation of the magnetic properties of the individual SRO layers \cite{Yang2021Sup}. Since only a single $T_C$ is observable, we attribute similar magnetic properties to each SRO layer in our heterostructure.

For the description of the magnetoresistance (MR) we develop the model presented in Eq. \ref{eq:MR}. Starting from two spin-polarized conduction channels comparable to the model of the anomalous Hall effect \cite{Thiel2020Sup}, we assume the total magnetoresistance as sum of the independent magnetoresistance contributions of these two channels.

The MR of each channel displays a quadratic dependence of the magnetic flux density $(B^2)$, which we express with the common relation $B\propto H+M$ with external magnetic field $H$ and magnetisation $M$ for ferromagnetic materials. The parameters $A$ and $C$ are fitting constants to take into account the different contribution strength of both channels.

The MR persist of three contributions: one displaying a purely quadratic field dependence ($H^2$), one showing a quadratic dependence of the magnetisations ($M^2_{s1}\left(H\right)$ and  $M^2_{s2}\left(H\right)$) and one with mixed contribution ($H\cdot M_{s1}\left(H\right)$ and  $H\cdot M_{s2}\left(H\right)$).

In Fig. \ref{fig-10K-60K-80K-oop-MR-seperated} we adapt our model to the temperature dependent MR measurements. The experimental data and the resulting simulation curve is displayed in Fig. \ref{fig-10K-60K-80K-oop-MR-seperated}(a), (c), (e), whereas we show the different contributions of the model in Fig. \ref{fig-10K-60K-80K-oop-MR-seperated}(b), (d) and (f). Here it becomes apparent that the MR is dominated by the mixed contribution (for example at $T=10$\,K or $T=80$\,K). At the temperature-dependent crossover the magnitude of the $H^2$ contribution and the mixed contribution have same magnitude but opposite sign.

Fig. \ref{fig-10K-AMR} shows the magnetoresistance at 10\,K for three different magnetic field orientations. For a field orientation perpendicular to the current flow ($H\parallel z$, $H\parallel y$) a positive MR is found, while a negative MR occurs when the magnetic field is parallel to the current ($H\parallel x$).

As shown in Fig. \ref{fig-T-dep-MR-Hall-K5-K10}, both Hall effect and magnetoresistance for heterostructures with different number of bilayer repetitions display rather similar behavior. This indicates, that our mayor conclusions on the physical properties of the [(SRO)$_5$/(SIO)$_2$)]$_{k}$ heterostructure, drawn on the $k=10$ samples, are independent on the number of repetitions of the (SRO)$_5$/(SIO)$_2$) bilayer structure.

\begin{widetext}
	\begin{eqnarray}
		\frac{\Delta\rho_{xx}}{\rho_{0}}&=& \Bigl(\frac{\Delta\rho_{xx}}{\rho_{0}}\Bigr)_{\text{s1}}+\Bigl(\frac{\Delta\rho_{xx}}{\rho_{0}}\Bigr)_{\text{s2}}\nonumber\\
		&\propto&  A\Bigl[H+M_{s1}\cdot\tanh\left(\frac{H-H_{c1}}{H_{s1}}\right)\Bigr]^2+C\Bigl[H+M_{s2}\cdot\tanh\left(\frac{H-H_{c2}}{H_{s2}}\right)\Bigr]^2\nonumber\\
		&\propto& 
		H^2 + \Bigl(M_{s1}\cdot\tanh\left(\frac{H-H_{c1}}{H_{s1}}\right)+ M_{s2}\cdot\tanh\left(\frac{H-H_{c2}}{H_{s2}}\right)\Bigr)\cdot H\nonumber\\
		& +&  \Bigl(M^2_{s1}\cdot\tanh^2\left(\frac{H-H_{c1}}{H_{s1}}\right)+ M^2_{s2}\cdot\tanh^2\left(\frac{H-H_{c2}}{H_{s2}}\right)\Bigr)\label{eq:MR}
	\end{eqnarray}
\end{widetext}

\end{document}